\documentclass[a4paper,11pt]{article}
\usepackage[utf8]{inputenc}
\pdfoutput=1
\usepackage{multirow}
\usepackage{jheppub}
\usepackage{subfig}
\usepackage{tikz}
\allowdisplaybreaks
\newcommand{\bmt}{\begin{pmatrix}}
\newcommand{\emt}{\end{pmatrix}}
\newcommand{\ba}{\begin{array}{c}}
\newcommand{\ea}{\end{array}}
\newcommand{\be}{\begin{equation}}
\newcommand{\ee}{\end{equation}}
\newcommand{\bea}{\begin{eqnarray}}
\newcommand{\eea}{\end{eqnarray}}

\newcommand{\bi}{\begin{itemize}}
\newcommand{\ei}{\end{itemize}}

\newcommand{\baz}{\begin{array}{cc}}
\newcommand{\mathsym}[1]{{}}

\newcommand{\bt}{\begin{tabular}}
\newcommand{\et}{\end{tabular}}

\newcommand{\benu}{\begin{enumerate}}
\newcommand{\eenu}{\end{enumerate}}
\def\m{\scriptstyle}
\def\l{\lambda}
\newcommand{\bav}{\begin{array}{cccc}}

\title{  Mitigating Direct Detection Bounds in  Non-minimal Higgs Portal Scalar Dark Matter Models}

\author[a] {Subhaditya Bhattacharya,}
\author[a]{Purusottam Ghosh,}
\author[b]{Tarak Nath Maity,}
\author[b,c]{Tirtha Sankar Ray}
\affiliation[a]{Department of Physics, Indian Institute of Technology Guwahati, North Guwahati, Assam- 781039, India}
\affiliation[b]{Department of Physics, Indian Institute of Technology Kharagpur, Kharagpur 721302, India}
\affiliation[c]{Centre for Theoretical Studies, Indian Institute of Technology Kharagpur, Kharagpur 721302, India}
\emailAdd{subhab@iitg.ernet.in}
\emailAdd{pghoshiitg@gmail.com}
\emailAdd{tarak.maity.physics@gmail.com}
\emailAdd{tirthasankar.ray@gmail.com}

\abstract{
The minimal   Higgs portal dark matter model is increasingly in tension with recent results form direct detection experiments like LUX  and XENON. In this paper we make a systematic study of simple extensions of the $ \mathbb{Z}_2$ stabilized singlet scalar Higgs portal scenario in terms of  their prospects at direct detection experiments. We consider  both enlarging the stabilizing symmetry to  $\mathbb{Z}_3$  and incorporating  multipartite features in the dark sector. We demonstrate that in these  non-minimal models the interplay of  annihilation, co-annihilation  and  semi-annihilation processes considerably relax  constraints from present and proposed direct detection experiments while simultaneously saturating observed  dark matter relic density. We explore  in particular the resonant semi-annihilation channel  within the  multipartite  $\mathbb{Z}_3$   framework  which results in new unexplored regions of parameter space that would be difficult to constrain by direct detection experiments in the near future.  The role of dark matter exchange processes within multi-component $\mathbb{Z}_3 \times \mathbb{Z}_3'$ framework is illustrated. We make quantitative  estimates to  elucidate the role of  various annihilation processes in the different  allowed regions of parameter space within these models. }


\begin{document}

\maketitle
\flushbottom

%%%%%%%%%%%%%%%%%%%%%%%%%%%%%%%%%%%%%%%%%%%
\section{Introduction}
\label{sec:intro}
%%%%%%%%%%%%%%%%%%%%%%%%%%%%%%%%%%%%%%%%%%%
Existence of dark matter (DM) is supported from many astrophysical evidences like rotation curve of galaxies  \cite{Rubin:1967msa, Rubin:1970zza}, anisotropies in CMBR \cite{Hu:2001bc}  and observations in bullet cluster \cite{Clowe:2006eq}. It is a possibility that DM is particulate and may even have some non-gravitational interaction with the Standard Model (SM) sector. This gives the  well discussed possibility  of DM composed of thermal relic of cosmologically stable particles \cite{Jungman:1995df, Munoz:2003gx, Bertone:2004pz, Bergstrom:2012fi, Arcadi:2017kky}.

The key feature of this paradigm is having a  cosmologically stable DM candidate. This is usually ensured in particle physics models by invoking some symmetry arguments. The $\mathbb{Z}_N$ discrete symmetries provide the simplest realization of this stabilizing symmetry and is commonly employed in extensions of the SM \cite{Batell:2010bp}. These symmetries can also arise as subgroup of  broken continuous symmetry groups. While the manifest discrete symmetry in the Lagrangian typically prevent any decay it still allows number changing processes between the dark sector and the SM. These number changing processes are crucial in maintaining the thermal equilibrium between the two sectors in the early stages of evolution of the Universe. Once the Universe starts expanding  these processes becomes less effective before finally  stalling, leading to the standard framework for  DM freeze out,  leaving behind  a  relic density  observable till the present epoch. The relic abundance has been precisely estimated from  CMBR studies at WMAP \cite{Komatsu:2010fb} and  then at PLANCK  \cite{Ade:2013zuv} experiments, to be in the range   $0.1133 \leq \Omega_{\rm DM} h^2 \leq 0.1189$.

The  simplest  number changing process that may be operative between the DM and SM sectors, allowed by the stabilizing symmetry, is the  so called DM annihilation. Where typically two DM particles annihilate to produce two or more SM states\footnote{For other non-standard annihilation possibilities see \cite{Dolgov:1980uu, Hochberg:2014dra, Dey:2016qgf}}.  This is effective in reducing the number density of the DM and lead to the freeze out. For weak scale mass and annihilation cross sections, these processes leads to the   Weakly Interacting Massive Particle (WIMP) framework \cite{Lee:1977ua}. However these same processes can be probed by the direct detection experiments \cite{Goodman:1984dc, Lewin}.

 Non observation of DM in direct detection experiments provides some of the most stringent bound on the DM models constraining  the annihilation processes as an effective mechanism to drive freeze out.  This correlation provides  the motivation for the  large number of direct detection  experiments that are in operation or have been proposed.
The Large Underground Xenon (LUX) experiment is   a dual-phase Xenon detector operating at the Sanford Underground Research Facility. First results of LUX \cite{Akerib:2013tjd} set forth a minimum upper limit on WIMP-nucleon  spin independent (SI) cross section of $7.6 \times 10^{-46} ~\rm cm^{2}$ at a WIMP mass of $33$ GeV.  XENON  is another experiment in operation at the Laboratori Nazionali del Gran Sasso, using ultra pure liquid Xenon as  WIMP  target. XENON100 experiment \cite{Aprile:2012nq} gathered data for 13 months between  2011 and 2012, reaching a minimum sensitivity of $2 \times 10^{-45} ~\rm cm^{2}$ at a DM mass of $55$ GeV with $90\%$ confidence level. The upgraded XENON1T  which acquired data for $34.2$ days have recently published first results \cite{Aprile:2017iyp}  already reaching sensitivity comparable to LUX and is  expected to  have increased  sensitivity in near future  with more data. Future proposals include the next generation XENONnT projected to achieve minimum spin-independent WIMP nucleon cross section $1.6 \times 10^{-48} ~\rm cm^{2}$ at WIMP mass of $50$  GeV  \cite{Aprile:2015uzo}. Dark matter WIMP search with liquid xenon (DARWIN) \cite{Aalbers:2016jon} will be an experiment for the direct detection of DM using multi-ton liquid xenon. This experiments can be sensitive to Spin-independent  DM-nucleon cross section of $2.5 \times 10^{-49} ~ \rm cm^{2}$ at a WIMP mass of $40$ GeV \cite{Aalbers:2016jon}.  Note that collider searches at  Large Hadron Collider (LHC)  puts a considerably  weaker bound on the intermediate mass (100-1000 GeV)  Higgs portal dark matter candidates compared to the direct detection constraints \cite{Arcadi:2017kky}.

These ongoing and proposed experiments mandates a closer look at various WIMP DM models and their prospects at these direct detection experiments. In this paper we will confine ourselves to a class of models where the  SM singlet  scalar dark sector communicates with the SM through a coupling with the Higgs. These, so called, Higgs portal models provides a simple framework for WIMP DM and are subject to extensive discussion in the literature \cite{Bertone:2004pz}.
Null results at direct detection experiments  \cite{Akerib:2013tjd, Aprile:2017iyp}  has already put a strong bound \cite{He:2016mls, He:2008qm} on the minimal Higgs portal models where a single scalar DM is stabilized by a discrete $ \mathbb{Z}_2$ symmetry.  Other than the extremely tuned Higgs mass pole region, the  well motivated \cite{Beniwal:2017eik, Artymowski:2016tme, Vaskonen:2016yiu, Cline:2017qpe} minimal  Higgs portal model has been pushed to a heavy DM mass region by XENON1T data \cite{Aprile:2017iyp}, which can be  further excluded by continued non observation in the immediate next generation experimental results.  In this article we make a systematic study of the simple extension of this framework that can  evade these constraints  while remaining a viable DM candidate.

 Possible augmentation  of the minimal model  can be done by enlarging the  stabilizing symmetry group from $\mathbb{Z}_2$ to $\mathbb{Z}_N$ or by introducing multi-particle  dark sector. These non-minimal models facilitate non-standard  number changing channels like semi-annihilation \cite{D'Eramo:2010ep} and  co-annihilation \cite{Edsjo:1997bg}, which  cannot be explored at direct detection experiments.  We find that within multipartite $\mathbb{Z}_2$ framework co-annihilation and   mediated annihilation processes  ameliorate some of the direct search bounds  \cite{Casas:2017jjg,  Ghorbani:2014gka}.  Further we investigate multipartite $\mathbb{Z}_3$ model  which have a significantly enriched DM phenomenology. Here the interplay of semi-annihilation and resonant semi-annihilation  together with co-annihilation and annihilation processes uncover a large region  of unexplored parameter space which satisfy both relic density and direct search bounds.   Finally we explore the DM exchange processes in two component DM scenario with a  $\mathbb{Z}_3 \times \mathbb{Z}_3'$  stabilizing symmetry. Interestingly, in certain regions of parameter space of this scenario  both  the DM states can be detected at the  next generation experiments.  We make an organized study of these frameworks taking each scenario one by one in increasing order of complexity. A detailed numerical scan of the parameter space is performed to explore the intricate interface of the various number changing process that are operative  in a given framework.

The rest of the paper is organized as follows. In section \ref{sec:status-z2} we present multi-particle $\mathbb{Z}_2$ framework.  In section \ref{sec:z3} we briefly  review the  impact of semi-annihilation in  scalar DM models stabilized by a $\mathbb{Z}_3$ symmetry,   before we present  a rigorous  study of multi-particle $\mathbb{Z}_3$ framework  including a detailed discussion of resonant semi-annihilation. And in section \ref{sec:teo-comp-z3}, we focus two component DM  under $\mathbb{Z}_3 \times \mathbb{Z}_3'$. In section \ref{nnnn} we briefly discuss tree level vaccum stability and unitarity constraints.  Finally we conclude in section \ref{sec:summary}.

%%%%%%%%%%%%%%%%%%%%%%%%%%%%%%%%%%%%%%%%%%
\section{Minimal $\mathbb{Z}_2$ Model and Extensions}
\label{sec:status-z2}
%%%%%%%%%%%%%%%%%%%%%%%%%%%%%%%%%%%%%%%%%%
The operators facilitating the  thermal freeze-out involve DM and SM fields which can be  decomposed into $\mathcal{O}_{SM-DM}\sim \mathcal{O}_{DM}\mathcal{O}_{SM},$ assuming SM sector do not transform under stabilizing symmetry of the dark sector. The simplest renormalizable operator of such kind can be written by ideating the existence of a real scalar singlet DM ($\phi$) interacting to SM through Higgs portal interactions as $\phi^2H^\dagger H$ and has been studied exhaustively in the literature \cite{Lerner:2009xg}. The Lagrangian can be written as,
\begin{eqnarray}
-\mathcal{L}_{\rm DM-Higgs} &=& 
- \mu_{H}^2 (H^{\dagger} H-\frac{v^2}{2}) 
+ \lambda_{H} (H^{\dagger} H-\frac{v^2}{2})^2 \nonumber \\
&&+ \frac{1}{2}  m_{\phi}^2 \phi^2
+ \frac{ \lambda_{s}}{4!} \phi^4
 + \frac{1}{2} \lambda_{h} \phi^2 (H^{\dagger} H-\frac{v^2}{2})  \,.
\label{eq:pot}
\end{eqnarray}
This setup  implies  the presence of an unbroken $\mathbb{Z}_2$ symmetry under which $\phi \to -\phi$ while all SM particles are even, making the field $\phi$ stable.   The relevant  annihilation processes that drive the freeze out  are depicted in Fig.~\ref{sf:annhz22}.  However, this minimal setup has been pushed to uncomfortable corner by recent results from direct search data in LUX 2016 \cite{Akerib:2016lao} and XENON1T \cite{Aprile:2017iyp}.      
The remaining  unconstrained region are now confined to the   tuned Higgs resonance region or for heavy DM mass $  m_{\phi} \gtrsim 500$ GeV. The heavy DM region lies just below the present direct detection constraints and  most of the allowed region will be explored in the next generation experiments.  Admitting a tuning of the DM mass  to be $ m_{\phi} \approx m_h/2 \sim 125/2$ GeV  allows considerable relaxation of the direct search bounds while still satisfying the relic density calculations.
Back of the envelope calculation show that  at the resonance  $\langle \sigma v \rangle \approx 24 ~ \lambda^2_h~ \rm GeV^{-2}$ assuming a Higgs width $\Gamma_h= 4 $ MeV \cite{Khachatryan:2016vau}. This  saturates the relic density bound  when $\langle \sigma v \rangle$ is equal to $\sim 0.1$ pb. Estimating the Higgs portal coupling from here we find that the direct detection cross section can be  as low as $1.69 \times 10^{-52}~ \rm cm^2$. Thus is expected to remain mostly  unconstrained by direct detection experiments in near future. 

 A permissible Higgs portal DM with intermediate mass necessarily imply extensions of the this framework. The simplest possible generalization to this framework is to add more singlet scalars states to the dark sector. If they transform under different symmetries, say $\mathbb{Z}_2^{'} \times \mathbb{Z}_2^{''} \cdots$ and so on, then all of them are stable leading to a multi-component DM framework.  The impact of the  DM-DM interactions in the two component framework under $\mathbb{Z}_2 \times \mathbb{Z}_2^{'}$  has been studied in \cite{Bhattacharya:2016ysw, Karam:2016rsz}.  The dark sector exchange interaction play the role of a \textit{see-saw} between the two components. While the lighter component  behave like the single component framework with early detection possibility, the heavier component may have suppressed direct detection cross section. This suppression for the heavier component  arises  when its contribution to the total DM relic abundance is relatively small.  When the DM masses and couplings to SM are same the symmetry is enlarged to $O(N)$ for $N$ component  DM scenario. The exchange process now vanishes and requires all the DM components to have larger annihilation cross-section and larger DM-SM couplings making them more constrained by direct detection experiments \cite{Drozd:2011aa}.

When the dark sector particles are charged under the same stabilizing symmetry, the heavier one can decay to the lighter component yielding a single component DM framework while the dark sector remains multipartite. The presences of extra particles open new non-standard number changing processes in the dark sector  like co-annihilation and mediated annihilation. As will be detailed in the rest of this section  this leads to considerable easing in many tensions  of the minimal model.

%%%%%%%%%%%%%%%%%%%%%%%%%%%%%%%%%%%%%%%%%%
\subsection{Two particles under \texorpdfstring{$\mathbb{Z}_2$}{Lg} : Co-annihilation \& Mediated annihilation}
\label{subsec:co-anni}
%%%%%%%%%%%%%%%%%%%%%%%%%%%%%%%%%%%%%%%%%%
We will consider the case where  the SM is augmented by two real scalar particles $\phi_1$ and $\phi_2$ odd under the same discrete $\mathbb{Z}_2$ symmetry and singlet under SM.  The relevant part of the Lagrangian is given by,
 \begin{eqnarray}
\label{potz22}
-\mathcal{L}_{DM-Higgs} &=&-\mu_{H}^2 (H^{\dagger}H-\frac{v^2}{2} )+{\ \lambda_H}(H^{\dagger}H-\frac{v^2}{2} )^2+\frac{1}{2}m_{\phi_1}^2 \phi_1^2+\frac{1}{2}m_{\phi_2}^2 \phi_2^2 \nonumber \\
&&+\frac{\lambda_{e_1}}{4}\phi_1^2\phi_2^2+\frac{\lambda_{e_2}}{3!}\phi_1^3\phi_2+\frac{\lambda_{e_3}}{3!}\phi_1\phi_2^3 + \frac{\lambda_{1s}}{4!}\phi_1^4+\frac{\lambda_{2s}}{4!}\phi_2^4\\ 
&&+\frac{1}{2}\lambda_{1h}\phi_1^2(H^{\dagger}H-\frac{v^2}{2} )+\frac{1}{2}\lambda_{2h}\phi_2^2 (H^{\dagger}H-\frac{v^2}{2} )+\lambda_{12h}\phi_1\phi_2 (H^{\dagger}H-\frac{v^2}{2} ).  \nonumber 
\end{eqnarray}
 Note that the coupling  $\lambda_{12h}$  arises as both of the components transform under same symmetry and give rise to the novel  co-annihilation and mediated annihilation channels depicted in Fig.~\ref{sf:coannhz22} and  Fig.~\ref{sf:newannsz22} respectively. Assuming without any loss of generality,  $m_{\phi_1}< m_{\phi_2}$  the lightest mass state  $\phi_1$ can be  identified as the potential DM candidate\footnote{In principle we can have a term like $m^2\phi_1\phi_2$ in the Lagrangian which is allowed by all symmetries of the theory. This can lead to a mass mixing between the states.  However,  note that the Lagrangian in Eq.~\ref{potz22} is written in  the mass basis  of $\phi_1$ and $\phi_2,$ assuming the mass matrix has been diagonalized. }. The $\phi_2$ state can promptly decay to $\phi_1,$ therefore do not effect the freeze out except through its contribution to the $\phi_1$ number changing process discussed above. This decay occurs through an off-shell Higgs and a schematic calculation can be found in Appendix  \ref{apnd:decay}.

%%%%%%%%%%%%%%%%%%
\subsubsection{Relic density }
\label{subsubsec:DMphz22}
\begin{figure}[t]

 \begin{center}\subfloat[\label{sf:annhz22}]{
    \begin{tikzpicture}[line width=0.5 pt, scale=1.1]
        %For phi,phi -> h,h point
       \draw[dashed] (-10,1)--(-9,0);
	\draw[dashed] (-10,-1)--(-9,0);
	\draw[dashed] (-9,0)--(-8.2,0);
	\draw[dashed] (-8.2,0)--(-7.2,1);
	\draw[dashed] (-8.2,0)--(-7.2,-1);
	\node  at (-10.4,-1) {$\phi_1$};
	\node at (-10.4,1) {$\phi_1$};
	\node [above] at (-8.6,0) {$h$};
	\node at (-6.9,1.2) {$h$};
	\node at (-6.9,-1.2) {$h$};
	\end{tikzpicture}
	\hspace{0.5cm}
%	% Forphi1 , phi1 -> phii, h
\begin{tikzpicture}[line width=0.5 pt, scale=1.1]
	\draw[dashed] (-6.4,1)--(-5.4,0);
	\draw[dashed] (-6.4,-1)--(-5.4,0);
	\draw[dashed] (-5.4,0)--(-4.6,0);
	\draw[dashed] (-4.6,0)--(-3.6,1);
	\draw[dashed] (-4.6,0)--(-3.6,-1);
	\node  at (-6.7,-1.2) {$\phi_1 $};
	\node at (-6.7,1) {$\phi_1$};
	\node [above] at (-5,0) {$h$};
	\node at (-3.4,1.2) {$\mbox{SM} $};
	\node at (-3.4,-1.2) {$\mbox{SM}$};
\end{tikzpicture}
	\hspace{0.5cm}
\begin{tikzpicture}[line width=0.5 pt, scale=1.1]
	\draw[dashed] (-2.5,1.5)--(-1.5,0.5);
	\draw[dashed] (-1.5,0.5)--(-1.5,-0.5);
	\draw[dashed] (-2.5,-1.5)--(-1.5,-0.5);
	\draw[dashed] (-1.5,0.5)--(-0.5,1.5);
	\draw[dashed] (-1.5,-0.5)--(-0.5,-1.5);
	\node  at (-2.7,1.5) {$\phi_{1}$};
	\node at (-0.3,1.5) {$h$};
	\node [right] at (-1.5,0.0) {$\phi_{1}$};
	\node at (-0.3,-1.5) {$h$};
	\node at (-2.7,-1.7) {$\phi_{1}$};     
      	 \end{tikzpicture}}
 \end{center}

 \begin{center}\subfloat[\label{sf:coannhz22}]{
     \begin{tikzpicture}[line width=0.5 pt, scale=1.1]
      %s channel  
        \draw[dashed] (-3,1)--(-2,0);
	\draw[dashed] (-3,-1)--(-2,0);
	\draw[dashed] (-2,0)--(-0.5,0);
	\draw[dashed] (-0.5,0)--(0.5,1);
	\draw[dashed] (-0.5,0)--(0.5,-1);
	\node  at (-3.4,-1) {$\phi_1$};
	\node at (-3.4,1) {$\phi_2$};
	\node [above] at (-1.25,0) {$h$};
	\node at (0.9,1) {$\rm SM$};
	\node at (0.9,-1) {$\rm SM$};
   \end{tikzpicture}}~~~~~~
   \subfloat[\label{sf:newannsz22}]{
     \begin{tikzpicture}[line width=0.5 pt, scale=1.1]
      %t channel
      \draw[dashed] (5,0)--(6,-1);
	\draw[dashed] (6,-1)--(6,-2);
	\draw[dashed] (5,-3.0)--(6,-2.0);
	\draw[dashed] (6,-1)--(7,0);
	\draw[dashed] (6,-2.0)--(7,-3.0);
	\node  at (4.6,0) {$\phi_{1}$};
	\node at (7.4,0) {$h$};
	\node [right] at (6,-1.5) {$\phi_{2}$};
	\node at (7.2,-3) {$h$};
	\node at (4.6,-3) {$\phi_{1}$};
  \end{tikzpicture}}
 \end{center}
 \caption{Processes contributing to relic density for two particles under $\mathbb{Z}_2$ framework (a) annihilation (b) co-annihilation (c) Mediated annihilation.}
 \label{fig:z22fd}
\end{figure}
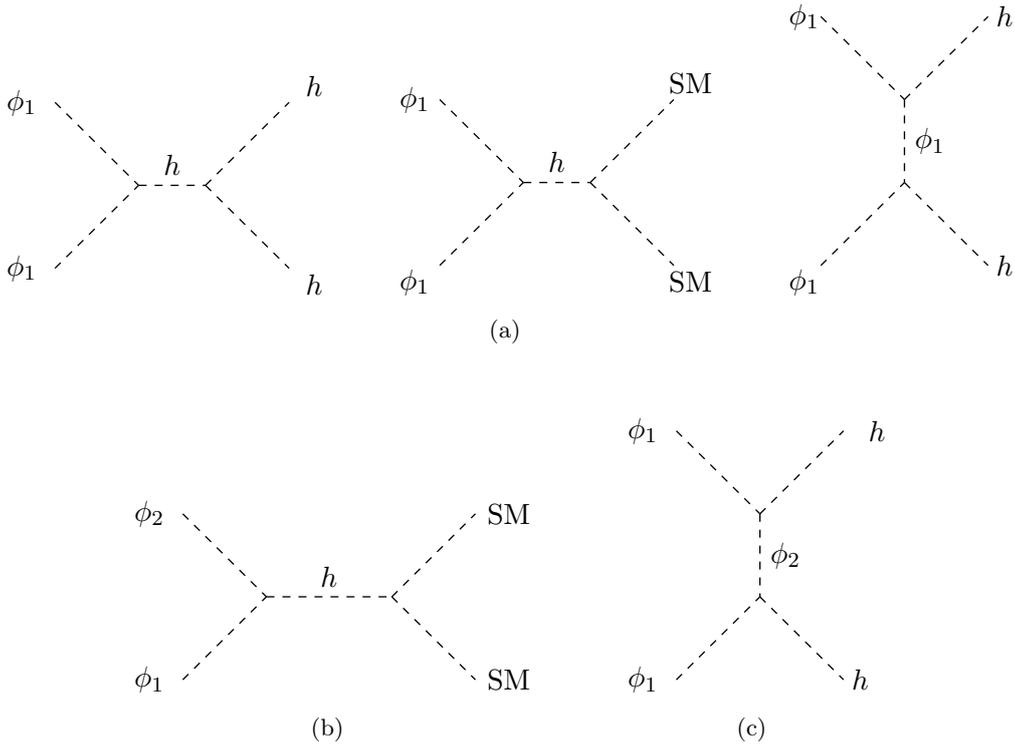

%%%%%%%%%%%%%%%%%%%%%%
The new set of  Feynman graphs corresponding to the number changing processes of DM are shown in Fig.~\ref{sf:coannhz22} and \ref{sf:newannsz22}. With a small annihilation cross section, as mandated by direct detection results, relic density bound can be saturated using the co-annihilation process depicted in Fig.~\ref{sf:coannhz22} if the masses of the two sates are relatively degenerate. However, for large mass gap  the  $t$-channel mediated annihilation process  in Fig.~\ref{sf:newannsz22} takes up a  major role in controlling DM relic density.  These novel  process are of interest because while they change the number density of the DM and thus aid freeze out they do not contribute to direct detection cross section. The co-annihilation channels do not contribute as the  mass gap between the co-annihilating  states kinematically forbid the corresponding direct detection process. The t-channel mediated annihilation with Higgs in the final state  couple to the nucleon at 1-loop. So remains relatively unconstrained by direct detection.

Presence of $\lambda_{e_i}$ in Eq.~\ref{potz22} essentially influences the evolution of the DM density in two ways: (i) From the decay of $\phi_2$, DM is produced and (ii) DM  scattering  processes: $\phi_2\phi_2 \to \phi_1\phi_1, ~\phi_2\phi_2 \to \phi_2 \phi_1,$ and $ \phi_2\phi_1 \to \phi_1 \phi_1.$  However, due to the prompt decay $\phi_2 \rightarrow \phi_1 X$ , these exchange processes have no role in setting the relic density for $\phi_1.$
Assuming that all the $\phi_2$ will ultimately transform themselves through decay processes to $\phi_1$ one can write down the Boltzmann equation for this  case in terms of the total DM relic density $n=\sum{n_{\phi_i}}=n_{\phi_1}+n_{\phi_2}$ as \cite{Edsjo:1997bg},
\begin{eqnarray}
\frac{dn}{dt}+3Hn&=& -\langle \sigma v \rangle_{eff}(n^2-{n^{eq}}^2)
\end{eqnarray}
where  
\be
\langle \sigma v \rangle_{eff}= \langle \sigma v \rangle_{\phi_1\phi_1\to SM}+ \langle \sigma v \rangle_{\phi_1\phi_2\to SM}(1+\frac{\Delta m}{m_1})^{3/2}e^{-\frac{\Delta m}{T}}~.
\nonumber
\ee
The relic density can easily be obtained approximately from the above equation as  $ \Omega h^2  \approx  (0.1~{\rm pb})/\langle \sigma v \rangle_{eff}$ \cite{Kolb:1990vq,Bhattacharya:2016ysw}. Note here that co-annihilation effect reduces with larger mass differences $\Delta m$ due to the Boltzmann suppression of $\exp (- \Delta m/T)$. In our numerical scans we will consider DM relic density to lie within:  $0.1133 \leq \Omega_{\rm DM} h^2 \leq 0.1189$ \cite{Ade:2013zuv}.

%%%%%%%%%%%%%%%%%
\subsubsection{Direct Detection }
\label{subsubsec:DDz22}
%%%%%%%%%%%%%%%%%%%
In this section we will discuss about direct search constraints on  multipartite $\mathbb{Z}_2$ model. In these experiments incoming DM flux  scatter with the nuclei in the  target crystals and the recoil can be searched for as a signal of the DM. Within the Higgs portal framework the  direct search of the DM  goes via $t-$ channel exchange of Higgs  as depicted in Fig.~\ref{fig:direct}. The spin independent direct search cross section of DM-Nucleon scattering reads \cite{Cline:2013gha},
  \begin{eqnarray} \label{eq:dd-sigma}
   \sigma_{n}^{SI}=\frac{\lambda_{1h}^2 f_n^2}{4\pi } \frac{\mu_n^2 m_n^2}{m_h^4m_{\phi_i}^2}~,
    \end{eqnarray}
 where $\mu_n= m_{n} m_{\phi_i}/(m_{n}+m_{\phi_i})$,  $m_n$ is the mass of the nucleon and {nucleon form factor, $f_n \approx 0.28$ \cite{Alarcon:2011zs,Alarcon:2012nr}.
As we are dealing with scalar DM and also we do not have any axial interaction term, therefore relevant bound comes from spin independent interaction cross section only.  We will consider limits from the recent LUX 2016 \cite{Akerib:2016lao} and XENON1T \cite{Aprile:2017iyp} data from non observation of DM in direct detection experiments and compare projected sensitivity in XENONnT\cite{Aprile:2015uzo} and DARWIN \cite{Aalbers:2016jon} experiments to validate the model.  Note that unless otherwise stated, throughout this paper  we use of {\tt micrOmegas} \cite{Belanger:2014vza}  to study the spin independent direct detection cross section.

\begin{figure}[t]
$$
 \includegraphics[scale=0.5]{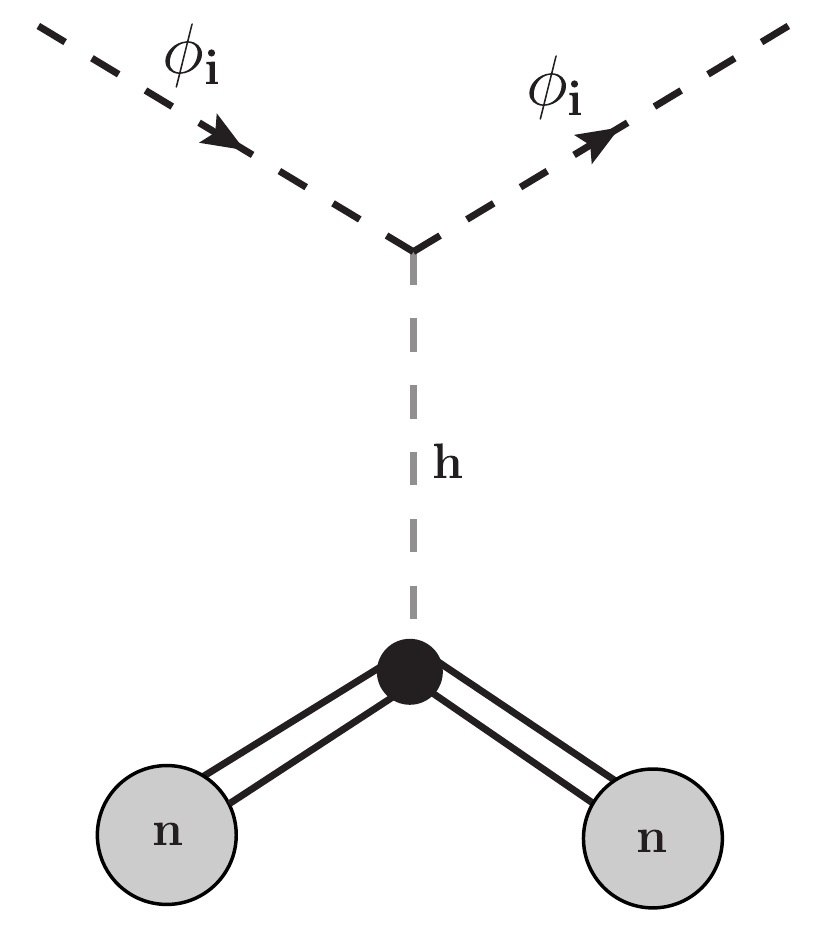}
  $$
 \caption{Feynman diagram for direct detection of scalar singlet DM.}
 \label{fig:direct}
\end{figure}

%%%%%%%%%%%%%%%%%%
\subsubsection{Numerical Scans and Analysis}
\label{subsubsec:DMphz22}
%%%%%%%%%%%%%%%%%%%%%%

The parameters of this model which govern DM phenomenology are essentially DM mass, mass of the co annihilating particle, their couplings to SM   i.e $\{m_{\phi_1}, m_{\phi_2}, \lambda_{1h}, \lambda_{2h}, \lambda_{12h}\} .$ We numerically scan the parameters of the model  to find the relic density allowed parameter space and then show the compatibility of the model with direct search experiments.  We utilize  {\tt micrOmegas} \cite{Belanger:2014vza}  to estimate both the relic density and the direct detection   spin independent cross sections as  summarized in Appendix \ref{apnd:mthd}. In the scans presented here the  parameter ranges  are chosen as follows,
\bea
1 ~{\rm GeV}< m_{\phi_1} < 1000~{\rm GeV},~2 \leq~\Delta{m}\equiv m_{\phi_2}- m_{\phi_1} \leq 1000~ \rm GeV, \nonumber\\
0.001 \leq\lambda_{1h}=\lambda_{2h}\leq 1,~0\leq \lambda_{12h}\leq 1 ~.\
\label{eqn:psz22}
\eea
\begin{figure}[t]
$$
 \includegraphics[scale=0.5]{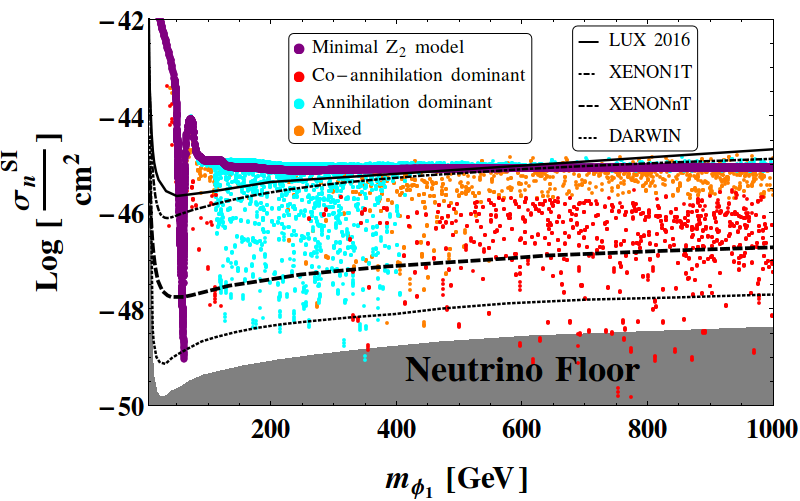}
$$
 \caption{Spin-independent DM-nucleon Direct Detection cross section for relic density allowed parameter space as function of DM mass for parameters given in Eq.~\ref{eqn:psz22} for two particles under $\mathbb{Z}_2$ model depicted in Eq.~\ref{potz22}. Co-annihilation dominant points are shown in red, annihilation dominant points are shown in cyan, mixed regions are shown by orange points. LUX 2016, XENON1T bound and XENONnT, DARWIN sensitivities are indicated. Shaded gray region represents Neutrino floor \cite{Billard:2013qya} for which direct search DM signal can not be distinguishable from background. } 
 \label{fig:DDz22}
\end{figure}

\begin{figure}[t]
\begin{center}
\centering
\subfloat[
\label{sf:m1m2xe1tz22} Relic density and XENON 1T allowed points are shown in  $m_{\phi_1}-m_{\phi_2}$  plane.]{
\includegraphics[scale=0.27]{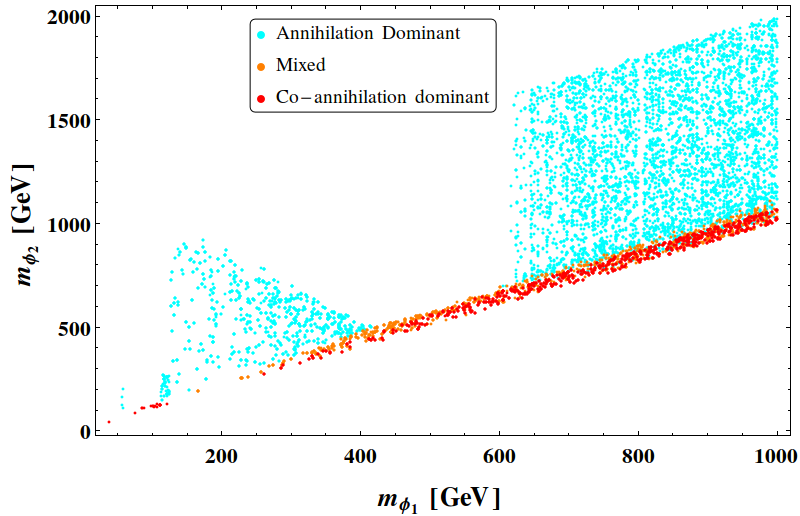}}
~~~
\subfloat[
\label{sf:lhm1z22} Relic density allowed points are shown in $m_{\phi_1}-\lambda_{1h}$ plane with direct search bounds.]{
\includegraphics[scale=0.27]{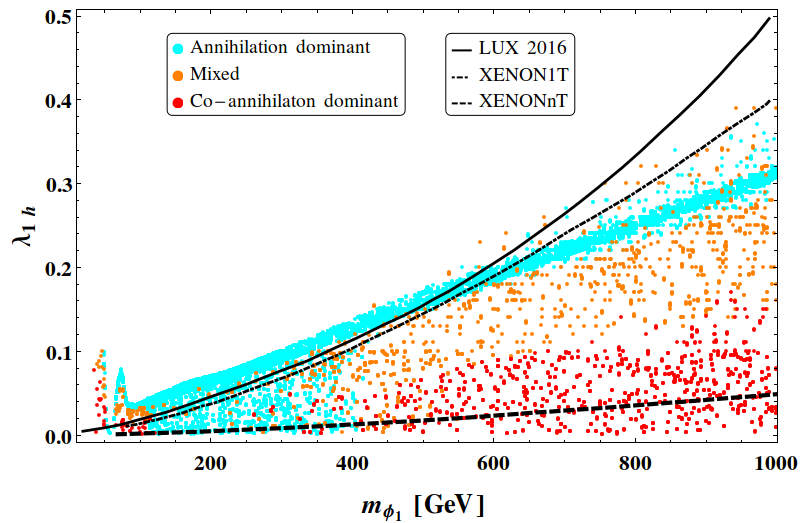}}
 \caption{Relic density and direct search allowed points for two particles under $\mathbb{Z}_2$ framework depicted in Eq.~\ref{potz22}. Co-annihilation dominant points are shown in red, annihilation dominant points are in cyan while the mixed regions are shown in orange.}
 \label{fig:ps-z2}
 \end{center}
\end{figure}

In Fig.~\ref{fig:DDz22} we have shown spin-independent direct detection cross section as function of DM mass ($m_{\phi_1}$) for the parameter space scanned. All the points in the plot satisfy the relic density constraint. The relic density allowed points are  further categorized  in terms of the dominant underlying number changing process that drives the freeze-out as, 
\begin{itemize}
\item Co-annihilation dominant (points in red)
\item Annihilation dominant (Cyan points)
\item Mixed  (Orange points)
\end{itemize}
If  the contribution of a particular process, co-annihilation or annihilation $\ge 80 \%$ to the relic density we assume that as dominant channel. For mixed cases, we choose all those points which are neither co-annihilation nor  annihilation is  dominant.  For the sake of comparison, we also depict relic density allowed parameter space points in minimal model, i.e. with one particle under $\mathbb{Z}_2$ in purple. The scan also indicates the background limit from solar, atmospheric and diffuse supernovae neutrinos in gray shaded region called neutrino floor, where detection of DM signal through direct search will be difficult \cite{Billard:2013qya}. 

\begin{figure}[t]
$$
 \includegraphics[scale=0.32]{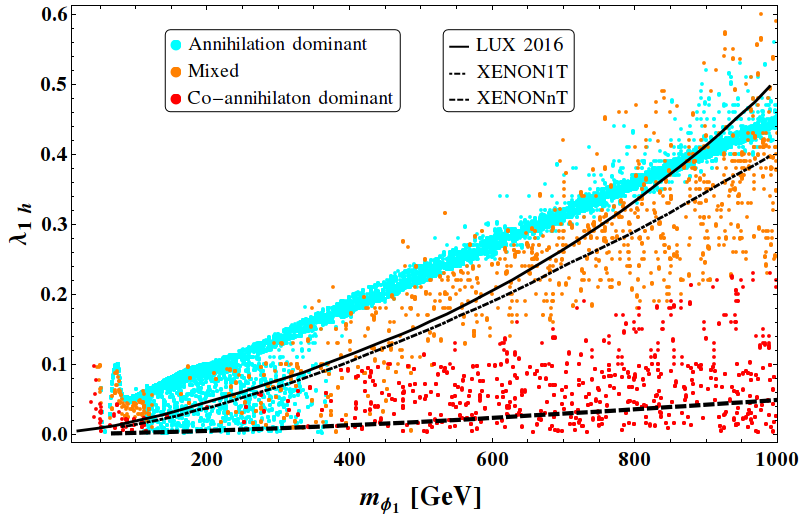}
$$
 \caption{Relic density allowed points when two complex scalar odd under same $\mathbb{Z}_2$ are shown in $m_{\phi_1}-\lambda_{1h}$ plane with direct search bounds.} 
 \label{fig:lhm1z22C}
\end{figure}

From the plot in Fig.~\ref{fig:DDz22}, we observe that points that satisfy relic density in this model easily survives the LUX 2016 \cite{Akerib:2016lao}, XENON1T \cite{Aprile:2017iyp} bound and can go beyond sensitivity of DARWIN,  hitting the neutrino floor. Thus, co-annihilation and mediated annihilation  not only resuscitate the intermediate mass scale of  $\mathbb{Z}_2$ Higgs portal scenario but in some regions of parameter space it remains unconstrained upto the projected limit of direct detection experiments. More interesting features arise when one investigates the underlying channels that contribute dominantly to the relic density calculations. For DM mass greater than $125~\rm GeV$ and below $400~\rm GeV$, dominant contribution to relic density and direct search allowed points come from mediated annihilation Fig.~\ref{sf:newannsz22}. However as we increase the mass of the DM consequently the NLSP  mass increases implying  progressively enhanced propagator suppression and the effect fades away at larger masses.  With larger DM mass, above $400~\rm GeV$,  for surviving points the dominant contribution  comes from  co-annihilation.  Admittedly this requires the NLSP to be relatively degenerate with $\Delta m/m_{\phi_1} \lesssim 20\%.$
The  relic density and direct search allowed points are plotted in $m_{\phi_1}-m_{\phi_2}$  plane in Fig.~\ref{sf:m1m2xe1tz22}. The  co-annihilation dominated points depicted in red predictably populate the region near $m_{\phi_2} \sim m_{\phi_1}$ due the Boltzmann factor. The first hump in the allowed parameter space for $m_{\phi_1}$ between $125 - 400$ GeV (cyan points in Fig.~\ref{sf:m1m2xe1tz22}) corresponds to mediated annihilation which fades out beyond $ m_{\phi_1} \sim 400$ GeV. The second allowed region for heavier masses $> 600$ GeV arises due to the traditional annihilation through the Higgs portal coupling of $\phi_1.$  The Fig. \ref{sf:lhm1z22} in the  $m_{\phi_1}-\lambda_{1h}$ plane clearly depicts that with large contribution from the co-annihilation and mediated annihilation the usual Higgs portal couplings can be suppressed. And as can be seen from Eq.~\ref{eq:dd-sigma} this effectively reduces the direct detection cross section.

Note that a complex scalar in $\mathbb{Z}_2$ framework essentially inherits two  degenerate degrees of freedom, here they will amount to two degenerate DMs. Operationally this will effectively yield : $\Omega_{\rm complex} = 2~\Omega_{\rm real}$ and the allowed parameter space can easily be scaled from above analysis. It has already been pointed out, that such two component model are relatively more constrained from direct search bounds. The allowed parameter space of this complex scalar scenario shown in Fig~\ref{fig:lhm1z22C}. However, in presence of co-annihilation  and mediated annihilation, some of these tensions eased.

%%%%%%%%%%%%%%%%%%%%%%%%%%%%%%%%%%%%%%%%%%%
\subsection{N-Scalar $\mathbb{Z}_2$ Model} \label{sec:mulz22}
%%%%%%%%%%%%%%%%%%%%%%%%%%%%%%%%%%%%%%%%%%%
One can easily extend  this framework by populating the dark sector with more than two real scalar singlet particles transforming under same $\mathbb{Z}_2$ symmetry. The  Lagrangian with $N$  such particles $\phi_i$ is given by,
 \begin{eqnarray}\label{eqn:potz2n}
& -&\mathcal{L}_{DM-Higgs} =-\mu_{H}^2 (H^{\dagger}H-\frac{v^2}{2} )+{\ \lambda_H}(H^{\dagger}H-\frac{v^2}{2} )^2+\frac{1}{2} \sum_{i=1}^{N} m_{\phi_{i}}^2 \phi^2_i  +\sum_{i=1}^{N}\frac{\lambda_{is}}{4!}\phi_i^4 \nonumber \\ 
&&+\sum_{\{i \neq j \neq k\}=1}^{N}\frac{\lambda_{ijk}}{2}\phi_i^2\phi_j\phi_k +\sum_{\{i \neq j\}=1}^{N}\frac{\lambda_{ij}}{4}\phi_i^2\phi_j^2+\sum_{\{i \neq j\}=1}^{N}\frac{\lambda_{ij}'}{3!}\phi_i^3\phi_j+\sum_{\{i \neq j \neq  k \neq l\}=1}^{N}\lambda_{ijkl}\phi_i\phi_j\phi_k\phi_l \nonumber \\
&&+\frac{1}{2} \sum_{i=1}^{N} \lambda_{ih}\phi_i^2(H^{\dagger}H-\frac{v^2}{2} )+\sum_{\{i\neq j\}=1}^{N} \lambda_{ijh}\phi_i\phi_j (H^{\dagger}H-\frac{v^2}{2} ).
\end{eqnarray}

\begin{figure}[t]
$$
\includegraphics[scale=0.45]{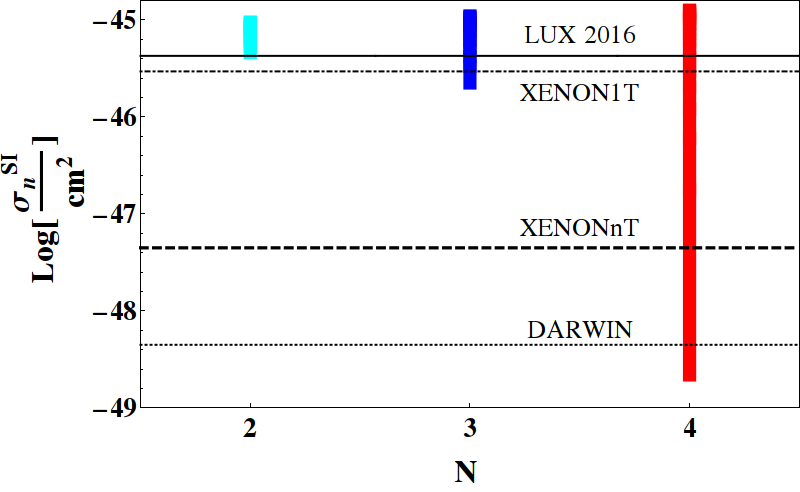}
$$
 \caption{Spin-independent DM-nucleon Direct Detection cross section for allowed relic density parameter space as a function of no. of scalar singlets ($N$) present with $\mathbb{Z}_2$ symmetry for $N=\{2,3,4\}$. The parameters chosen for the scan is indicated in Eq.~\ref{eqn:psz2n}. We also show the present bound from LUX 2016 and XENON1T sensitivities of XENONnT and DARWIN.}
 \label{fig:DDz2n}
\end{figure}
 
The lightest $\mathbb{Z}_2$ odd particle will be cosmologically stable and thus a DM candidate. Due to additional states in the odd sector we now have multiple copies of the co-annihilation and mediated annihilation channels assisting the freeze out of the DM.  It is then easy to appreciate that the limit on each  parameter gets much more relaxed compared to the two component case to survive the direct search bound after satisfying relic density.  A complete analysis of the model is computationally expensive and does not introduce any novel feature in the general discussion of non-standard number changing mechanism of the DM.  As an illustration of the impact of additional states we  perform a simplified  scan: 
 \begin{equation}
 \label{eqn:psz2n}
  m_{\phi_1}=200~\rm GeV, \Delta m_i \equiv m_{\phi_i}-m_{\phi_1} =10~\rm GeV,~ 0.001 \leq\lambda_{1h}=\lambda_{ih}\leq 0.1,~  \lambda_{1ih} \leq 0.112
 \end{equation}
 where $\phi_1$ is assumed to be the DM.  All other interactions between the DM states  have been put to zero. Mass difference is deliberately kept low to have appreciable contribution from co-annihilation. The direct search cross-section for relic density allowed points are shown in Fig.~\ref{fig:DDz2n} as a function of  $N$  that denote the number of scalar sates in the $\mathbb{Z}_2$ odd sector including the DM candidate. We can see that, given  the choice of parameters with relatively smaller values  of $\lambda_{12h}$ in Eq.~\ref{eqn:psz2n},  the  $N=2$ case can not satisfy LUX direct search bounds, whereas $N=4$ scenario can go upto the  DARWIN limit \footnote{ This should be contrasted with  a multi-component DM scenario with multiple particles that are odd under different 
 $\mathbb{Z}_2 \times  \mathbb{Z}_2' \times \mathbb{Z}_2'' \cdots  $.  In this case the scenario is expected to more constrained from direct detection bounds, as the number of such DM states are increased. }.

 %%%%%%%%%%%%%%%%%%%%%%%%%%%%%%%%%%%%%%%%%%%
\section{Minimal  \texorpdfstring{$\mathbb{Z}_3$}{Lg} Model and Extensions: Semi-Annihilation}
\label{sec:z3}
%%%%%%%%%%%%%%%%%%%%%%%%%%%%%%%%%%%%%%%%%%%

A complementary approach that can accommodate  processes which contribute to the DM decoupling but are not bounded by direct detection experiment is to enlarge the stabilizing symmetry $\mathbb{Z}_2 \rightarrow \mathbb{Z}_N.$ In this section we will consider the scenario where a discrete $\mathbb{Z}_3$ symmetry stabilizes  the dark sector \footnote{For a  study of $\mathbb{Z}_4$ models   including  comparison with the  minimal $\mathbb{Z}_3,$   see \cite{Belanger:2014bga}.}. A novel feature of this framework is the existence of the semi-annihilation processes that can potentially lead to relaxation in the direct detection bounds within Higgs portal models  \cite{Belanger:2012zr, Karam:2015jta}.

A  Higgs portal model with a single  SM singlet complex scalar  stabilized by a  $\mathbb{Z}_3$ has been discussed in \cite{Belanger:2012zr}. In this case, one requires necessarily a complex scalar field $\phi_1$ which transforms  non-trivially under $\mathbb{Z}_3$ as $\phi_1 \rightarrow \omega^n \phi_1$ where $\omega=\exp(i2\pi/3)$ and $n=1,2$ . Then invariant Lagrangian is given by, 
\begin{eqnarray} \label{eq:mz3}
-{\mathcal{L}}_{DM-Higgs} &=&-\mu_{H}^2 (H^{\dagger}H-\frac{v^2}{2} )+{\ \lambda_H}(H^{\dagger}H-\frac{v^2}{2} )^2+m^{2}_{\phi _{1}} \phi^{*}_{1}\phi _{1} +\frac{\mu_{1}}{3!} (\phi^{3}_{1}+\mbox{h.c})  \nonumber   \\
&+& \lambda_{1s}(\phi^{*}_{1}\phi _{1})^{2} 
+\lambda_{1h}(\phi^{*}_{1}\phi _{1})(H^{\dagger}H-\frac{v^2}{2} ).
 \end{eqnarray}
 \begin{figure}[htb!]
 \begin{center}
     \begin{tikzpicture}[line width=0.5 pt, scale=1.1]
       %s channel  
        \draw[dashed] (-1,1)--(0,0);
	\draw[dashed] (-1,-1)--(0,0);
	\draw[dashed] (0,0)--(1.5,0);
	\draw[dashed] (1.5,0)--(2.5,1);
	\draw[dashed] (1.5,0)--(2.5,-1);
	\node  at (-1.4,-1) {$\phi_{1}$};
	\node at (-1.4,1) {$\phi_{1}$};
	\node [above] at (0.75,0) {$\phi_{1}$};
	\node at (2.9,1) {$h$};
	\node at (2.9,-1) {$\phi^{*}_{1}$};
      %t channel
      \draw[dashed] (5,1)--(6,0);
	\draw[dashed] (6,0)--(6,-1.0);
	\draw[dashed] (5,-2.0)--(6,-1.0);
	\draw[dashed] (6,0)--(7,1);
	\draw[dashed] (6,-1.0)--(7,-2.0);
	\node  at (4.6,1) {$\phi_{1}$};
	\node at (7.4,1) {$\phi_{1}$};
	\node [right] at (6,-0.5) {$\phi_{1}$};
	\node at (7.2,-2) {$h$};
	\node at (4.6,-2) {$\phi^{*}_{1}$};
     \end{tikzpicture}
\end{center}
     \caption{Feynman diagrams for Semi-annihilation in $\mathbb{Z}_3$ model as in Eq.~\ref{eq:mz3}. }
     \label{fig:semi-anni}
\end{figure}
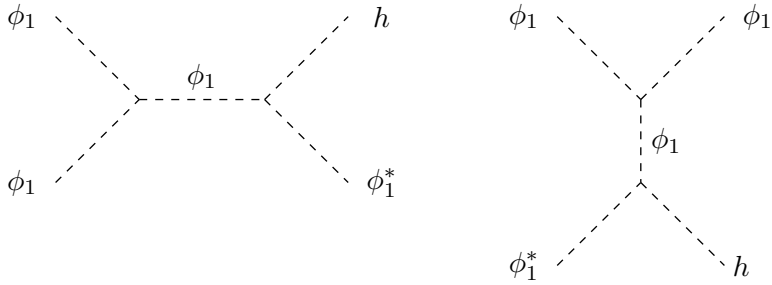
 The novel features in this framework arises from the $\phi^3_1$ term proportional to the dimensionful coupling $\mu_1.$ This leads to  DM to semi-annihilate by $\phi_1\phi_1 \to \phi_1 h$ through a  s-channel and a t-channel processes shown in Fig.~\ref{fig:semi-anni}. These are the new number changing channels that are now available in addition to the usual processes shown in Fig.~\ref{sf:annhz22} which are common to all Higgs portal models. The semi-annihilation process  is  operative  when  the  DM mass  becomes  heavier than Higgs mass $(m_{\phi_1}>m_h)$. Note that  its contribution reduces with increasing  DM mass  because of propagator suppression.

 \begin{figure}[t]
$$
 \includegraphics[scale=0.5]{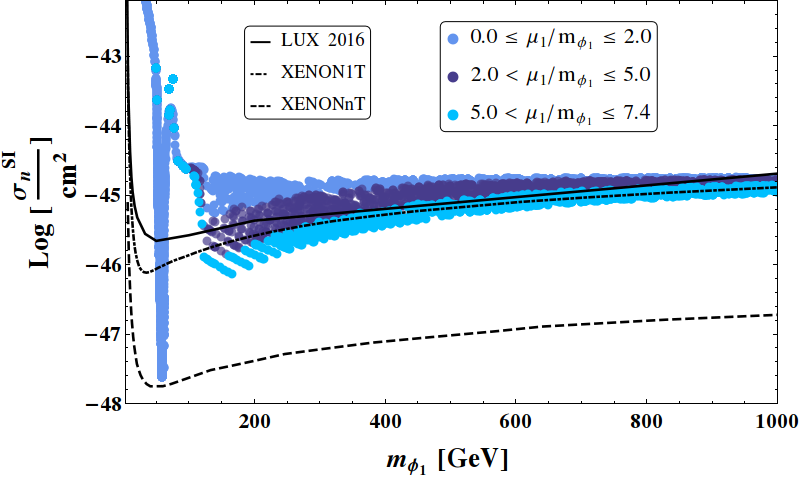}
$$
 \caption{Spin-independent DM-nucleon Direct Detection cross section of minimal $\mathbb{Z}_3$ model for allowed relic density parameter space as function of DM mass for parameter space dictated in Eq. \ref{eqn:psz3}. Three different colors represents three different  ranges  of $\mu_1.$} 
 \label{fig:RZ3a}
\end{figure}

The Boltzmann equation for relic density in presence of semi-annihilation is given by \cite{Belanger:2012zr},
\be
\frac{dn_{\phi_1}}{dt}+3Hn_{\phi_1}=-\langle \sigma v \rangle_{\phi_1\phi_1\to SM}(n_{\phi_1}^2-{n_{\phi_1}^{eq}}^2)- \frac{1}{2}\langle \sigma v \rangle_{\phi_1\phi_1\to \phi_1 SM}(n_{\phi_1}^2-n_{\phi_1}n_{\phi_1}^{eq})~,
\label{z3beq}
\ee 
where semi-annihilation is present in addition to the annihilation cross-sections. Semi-annihilation cross-section crucially depends on the dimensionful coupling $\mu_1$ and also the SM-DM coupling $\lambda_{1h}$. The pronounced effect from semi-annihilation is around lower DM mass ($ m_h \lesssim m_{\phi_1} \lesssim 400$) GeV where the propagator suppression is minimal. The right relic density can now be achieved for smaller values of the Higgs portal coupling ($\lambda_{1h}$) because of the assistance from the new channels. This improves the direct detection prognosis for the $\mathbb{Z}_3$ model. 

To illustrate the essential feature of this framework we perform a  three dimensional scan of the parameters  as follows,
\begin{equation}
50 ~{\rm GeV} \leq m_{\phi_1} \leq 1000~{\rm GeV},~0.001 \leq \lambda_{1h} \leq 1.0,~ \mu_1 \leq 7.4 m_{\phi_1}
\label{eqn:psz3}
\end{equation}
where we have set the upper limit on $\mu_1$ from vacuum stability considerations as detaile in section~\ref{nnnn}.
We follow the  methodology outlined in Appendix \ref{apnd:mthd}. The results of the scan is shown in DM mass versus direct search cross-section plane in Fig.~\ref{fig:RZ3a}. We show different choices of semi-annihilation parameter $\mu_1$ in different colors and it is obvious that the larger the $\mu_1$ is, for example, when we choose $5< \mu_1/ m_{\phi_1} \le 7.4$, we can reach the maximum sensitivity of the model by lowering spin independent cross section to nucleons while still saturating the relic limits. However, this  is still not sufficient to survive the XENONnT \cite{Aprile:2015uzo} except in the Higgs resonance region where the cross section can be  as low as $10^{-52}~ \rm{cm}^2$  as sketched previously. The  apparent lower limit  along the Higgs resonance branch  as depicted in Fig.~\ref{fig:RZ3a} is a result of our choices of the scan parameters.
 
%%%%%%%%%%%%%%%%%%%%%%%%%%%%%%%%%%%%%%%%%%%
\subsection{Two particles under \texorpdfstring{$\mathbb{Z}_3$}{Lg} : Resonant semi-annihilation}
\label{subsec:semi-and-co}
%%%%%%%%%%%%%%%%%%%%%%%%%%%%%%%%%%%%%%%%%%%
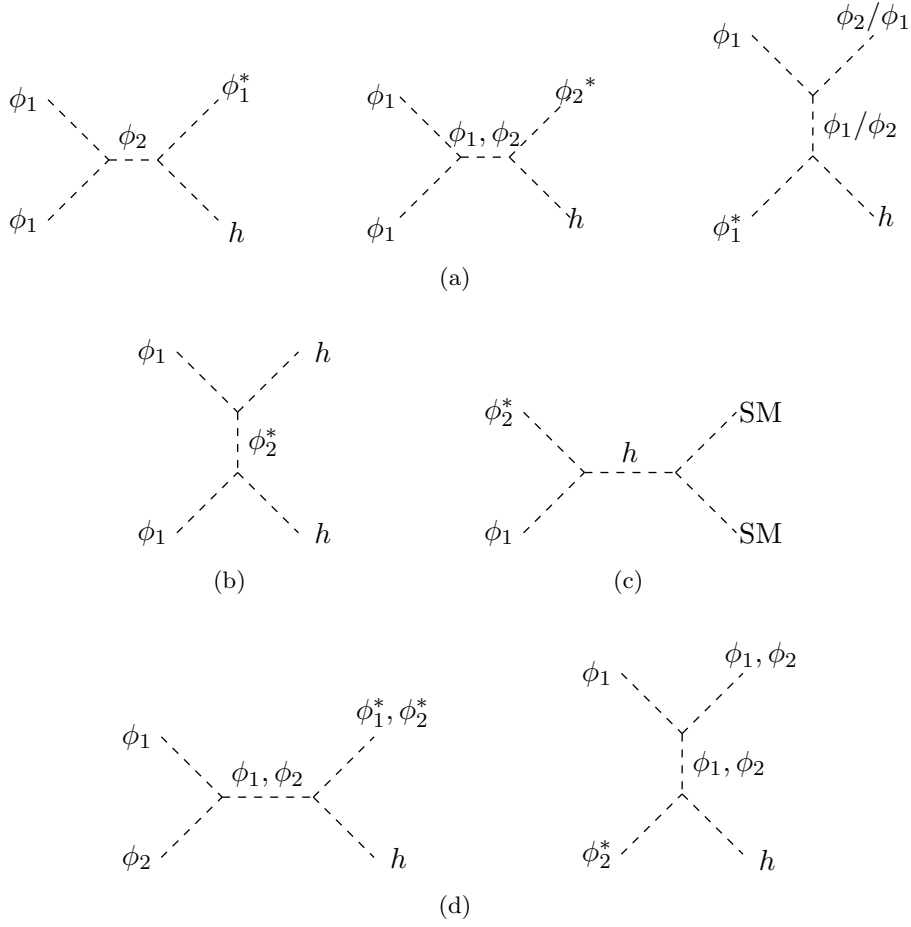
\begin{figure}[t] 
  \begin{center}
 \subfloat[\label{sf:semiannz32}]{ 
    \begin{tikzpicture}[line width=0.5 pt, scale=0.8]
	%For phi1 ,phi1 -> phii, h
%S-channel	
%
%      
       \draw[dashed] (-10,1)--(-9,0);
	\draw[dashed] (-10,-1)--(-9,0);
	\draw[dashed] (-9,0)--(-8.2,0);
	\draw[dashed] (-8.2,0)--(-7.2,1);
	\draw[dashed] (-8.2,0)--(-7.2,-1);
	\node  at (-10.4,-1) {$\phi_1$};
	\node at (-10.4,1) {$\phi_1$};
	\node [above] at (-8.6,0) {$\phi_2$};
	\node at (-6.9,1.2) {$\phi^*_1$};
	\node at (-6.9,-1.2) {$h$};
	\end{tikzpicture}
	\hspace{1cm}
%	% Forphi1 , phi1 -> phii, h
\begin{tikzpicture}[line width=0.5 pt, scale=0.8]
	\draw[dashed] (-6.4,1)--(-5.4,0);
	\draw[dashed] (-6.4,-1)--(-5.4,0);
	\draw[dashed] (-5.4,0)--(-4.6,0);
	\draw[dashed] (-4.6,0)--(-3.6,1);
	\draw[dashed] (-4.6,0)--(-3.6,-1);
	\node  at (-6.7,-1.2) {$\phi_1 $};
	\node at (-6.7,1) {$\phi_1$};
	\node [above] at (-5,0) {$\phi_1, \phi_2$};
	\node at (-3.5,1.1) {${\phi_2}^* $};
	\node at (-3.5,-1.0) {$h$};
\end{tikzpicture}
	\hspace{1cm}
\begin{tikzpicture}[line width=0.5 pt, scale=0.8]
	\draw[dashed] (-2.5,1.5)--(-1.5,0.5);
	\draw[dashed] (-1.5,0.5)--(-1.5,-0.5);
	\draw[dashed] (-2.5,-1.5)--(-1.5,-0.5);
	\draw[dashed] (-1.5,0.5)--(-0.5,1.5);
	\draw[dashed] (-1.5,-0.5)--(-0.5,-1.5);
	\node  at (-2.9,1.5) {$\phi_{1}$};
	\node at (-0.5,1.8) {$\phi_{2} / \phi_{1}$};
	\node [right] at (-1.5,0.0) {$\phi_{1}/ \phi_{2}$};
	\node at (-0.3,-1.5) {$h$};
	\node at (-2.9,-1.7) {$\phi^*_{1}$};     
      	 \end{tikzpicture}}
 \end{center}

  \begin{center}
   \subfloat[\label{sf:newannsz32}]{
     \begin{tikzpicture}[line width=0.5 pt, scale=0.8]
      %t channel 
      \draw[dashed] (-5,1)--(-4,0);
	\draw[dashed] (-4,0)--(-4,-1.0);
	\draw[dashed] (-5,-2.0)--(-4,-1.0);
	\draw[dashed] (-4,0)--(-3,1);
	\draw[dashed] (-4,-1.0)--(-3,-2.0);
	\node  at (-5.4,1) {$\phi_{1}$};
	\node at (-2.6,1) {$h$};
	\node [right] at (-4,-0.5) {$\phi^*_{2}$};
	\node at (-2.6,-2) {$h$};
	\node at (-5.4,-2) {$\phi_{1}$};
     \end{tikzpicture}} \hspace{1.5cm}
 \subfloat[\label{sf:coannhz32}]{
     \begin{tikzpicture}[line width=0.5 pt, scale=0.8]
      %s channel 

        \draw[dashed] (-1,1)--(0,0);
	\draw[dashed] (-1,-1)--(0,0);
	\draw[dashed] (0,0)--(1.5,0);
	\draw[dashed] (1.5,0)--(2.5,1);
	\draw[dashed] (1.5,0)--(2.5,-1);
	\node  at (-1.4,-1) {$\phi_1$};
	\node at (-1.4,1) {$\phi^*_2$};
	\node [above] at (0.75,0) {$h$};
	\node at (2.9,1) {$\rm SM$};
	\node at (2.9,-1) {$\rm SM$};
     \end{tikzpicture}}
 \end{center}
 
 \begin{center}\subfloat[\label{sf:cosemiannhz32}]{
     \begin{tikzpicture}[line width=0.5 pt, scale=0.8]
      %s channel  
        \draw[dashed] (-1,1)--(0,0);
	\draw[dashed] (-1,-1)--(0,0);
	\draw[dashed] (0,0)--(1.5,0);
	\draw[dashed] (1.5,0)--(2.5,1);
	\draw[dashed] (1.5,0)--(2.5,-1);
	\node  at (-1.4,-1) {$\phi_2$};
	\node at (-1.4,1) {$\phi_1$};
	\node [above] at (0.75,0) {$\phi_1,\phi_2$};
	\node at (2.8,1.4) {$\phi^*_1,\phi^*_2$};
	\node at (2.9,-1) {$h$};
     \end{tikzpicture}
     \hspace{1.5cm}
 %\subfloat[\label{sf:newannsz32}]{
     \begin{tikzpicture}[line width=0.5 pt, scale=0.8]
      %t channel
      \draw[dashed] (5,1)--(6,0);
	\draw[dashed] (6,0)--(6,-1.0);
	\draw[dashed] (5,-2.0)--(6,-1.0);
	\draw[dashed] (6,0)--(7,1);
	\draw[dashed] (6,-1.0)--(7,-2.0);
	\node  at (4.6,1) {$\phi_{1}$};
	\node at (7.3,1.3) {$\phi_1,\phi_2$};
	\node [right] at (6,-0.5) {$\phi_1,\phi_2$};
	\node at (7.4,-2.1) {$h$};
	\node at (4.6,-2) {$\phi^*_{2}$};
     \end{tikzpicture}}
 \end{center}
 
\caption{New Feynman diagrams in two particles under $\mathbb{Z}_3$ model {Eq.~\ref{eq:potz32}}  through (a) additional semi-annihilation (the left most diagram can exhibit resonant behavior) (b) Mediated annihilation (c) Co-annihilation (d) Novel co-annihilation. }
\label{fig:semiann_dig}
\end{figure}

We now turn to multipartite model by introducing  two complex scalar  $ \phi _{1}$  and $\phi _{2}$ charged under the same $\mathbb{Z}_{3}$ symmetry. The  Lagrangian is an generalization of Eq.~\ref{z3beq} and can be written as,
 \begin{eqnarray}
 \label{eq:potz32}
&-&{\mathcal{L}}_{DM-Higgs}=-\mu_{H}^2 (H^{\dagger}H-\frac{v^2}{2} )+{\ \lambda_H}(H^{\dagger}H-\frac{v^2}{2} )^2+m^{2}_{\phi_{1}} \phi^{*}_{1}\phi _{1}+m^{2}_{\phi _{2}} \phi^{*}_{2}\phi _{2}\nonumber  \\
 &+&\frac{\mu_{1}}{3!} (\phi^{3}_{1}+\mbox{h.c})+\frac{\mu_{2}}{3!} (\phi^{3}_{2}+\mbox{h.c})+\frac{\mu_{12}}{2!} (\phi^{2}_{1}\phi_2+\phi_2^2\phi_1+\mbox{h.c}) \\
 &+& \lambda_{1s}(\phi^{*}_{1}\phi _{1})^2+\lambda_{2s}(\phi^{*}_{2}\phi _{2})^2 +\lambda_{e}[(\phi^{*}_{1}\phi _{1})(\phi^{*}_{2}\phi _{2})+\{(\phi^{*}_{1}\phi _{2})^2+\mbox{h.c}\}] \nonumber \\
 &+& \lambda_{1h}(\phi^{*}_{1}\phi _{1})(H^{\dagger}H-\frac{v^2}{2} )+\lambda_{2h}(\phi^{*}_{2}\phi _{2})(H^{\dagger}H-\frac{v^2}{2} )+\lambda_{12h}[\phi^{*}_{1}\phi _{2}+\mbox{h.c}](H^{\dagger}H-\frac{v^2}{2})~.\nonumber
\end{eqnarray}
The $\mathbb{Z}_3$ symmetry group elements are $\{1,~\omega,~ \omega^2\}$ and the Lagrangian above is invariant if the dark sector particles $(\phi_1,~\phi_2)$ have identical  charges. If the charges are different the Lagrangian is modified, however, the essential features remains identical in both cases.  Without loosing any essential physics we will assume that  $m_{\phi_{1}} < m_{\phi_2} $ making $\phi_1$ as the DM candidate. 

%%%%%%%%%%%%%%%%%%%%%%%%%%%%%%%
\subsubsection{Relic Density}
\label{subsubsec:dmphz32}
%%%%%%%%%%%%%%%%%%%%%%%%%%%%%%%%
The lightest particle transforming under $\mathbb{Z}_3$ will be stable and serve as the DM candidate of the model, while the heavier one will have prompt decay to the DM. In this analysis, we will assume $\phi_{1}$ as the lightest stable particle and DM while $\phi_{2}$ is the next to lightest stable particle (NLSP). For further simplification we will  assume $\lambda_{1h}=\lambda_{2h}$. The Boltzmann equation governing the freeze-out of this DM will be given by:
\bea
\frac{dn_{\phi_1}}{dt}+3Hn_{\phi_1}&=&-\langle \sigma v \rangle_{\phi_1\phi_1\to SM}(n_{\phi_1}^2-{n_{\phi_1}^{eq}}^2)-\langle \sigma v \rangle_{\phi_1\phi_2\to SM}(n_{\phi_1}n_{\phi_2}-n_{\phi_1}^{eq}n_{\phi_2}^{eq}) \nonumber \\ 
&-&\frac{1}{2}\langle \sigma v \rangle_{\phi_1\phi_1\to\phi_1 SM}(n_{\phi_1}^2-n_{\phi_1}n_{\phi_1}^{eq})-\frac{1}{2}\langle \sigma v \rangle_{\phi_1\phi_1\to\phi_2 SM}(n_{\phi_1}^2-\frac{{n_{\phi_1}^{eq}}^2}{n_{\phi_2}^{eq}} n_{\phi_2})\nonumber \\
&-&\frac{1}{2}\langle \sigma v \rangle_{\phi_1\phi_2\to\phi_2 SM}(n_{\phi_1}n_{\phi_2}-n_{\phi_1}^{eq}n_{\phi_2}),
\eea 
where contributions from both semi-annihilation and co-annihilation dictates the thermal freeze-out of the DM on top of the annihilation cross-section of the DM ($\phi_1$).              
 The late time relic density for  $\phi_1$  depend on the number densities  $n_{\phi_1}$ and $n_{\phi_2}$ at the instance  of freeze out of $\phi_1,$ since any residual  $\phi_2$ eventually decay to $\phi_1.$  An estimation of this requires the simultaneous solution of the coupled Boltzmann equations for the two concerned species, however in our numerical results  we consistently utilize {\tt micrOmegas}. Note that  the bound on the values of $\mu$'s comes from vacuum stability considerations and  we  adopt a conservative choice of $\mu \leq 2 m_{\phi_1}$. Additional processes that contribute to the freeze out of the DM are shown in Fig.~\ref{fig:semiann_dig}: (i) additional semi-annihilation channels $\phi_1, \phi_1 \to \phi_{1,2}, h$ are   shown in Fig.~\ref{sf:semiannz32}, (ii) mediated annihilation shown in Fig.~\ref{sf:newannsz32}, (iii) Co-annihilation channels shown in Fig.~\ref{sf:coannhz32} and (iv) some novel  co-annihilation channels  that have numerically suppressed contribution, are  shown in Fig.~\ref{sf:cosemiannhz32}. All these processes while crucially assisting freeze out do not contribute to the  tree level DM-nucleon coupling, hence remains unconstrained from direct detection experiments.

%%%%%%%%%%%%%%%%%%%%%
\subsubsection{Resonant Semi-Annihilation}
\label{subsec:ressemi}
%%%%%%%%%%%%%%%%%%%%%
\begin{figure}[t]
\begin{center}
\subfloat[ \label{fig:m2omega} Resonance effect in semi-annihilation depicted in $m_{\phi_2}$ - $\Omega h^2$ plane for $m_{\phi_1}=300~ \rm GeV$. Other parameters are mentioned in Eq.~\ref{eq:ps-rsa}.]{
\includegraphics[scale=0.25]{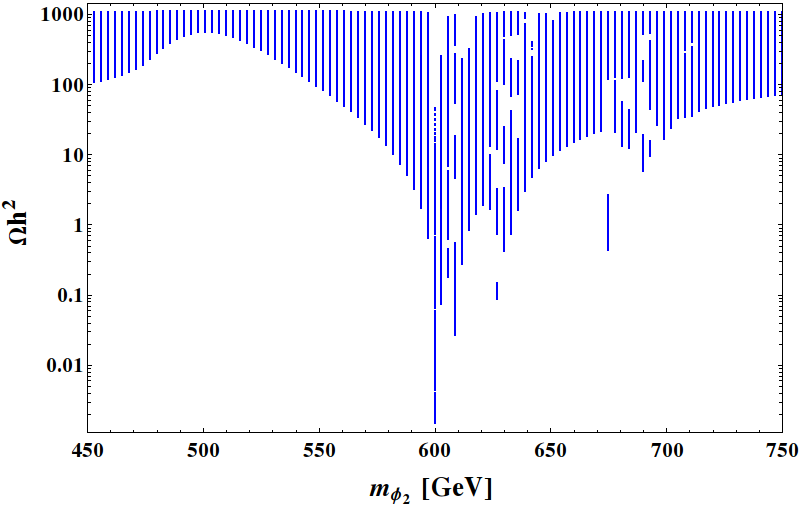}}
~~~
\subfloat[ \label{fig:m1omega} Resonance effect in $m_{\phi_1}$ - $\Omega h^2$ plane. Black thick line shows correct relic density.]{
\includegraphics[scale=0.25]{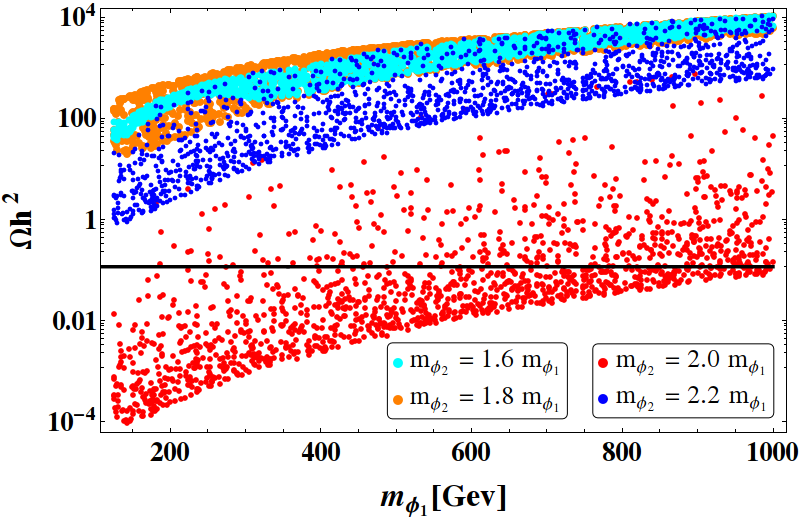}}
\caption{Resonant semi-annihilation in two particles under $\mathbb{Z}_3$ model Eq.~\ref{eq:potz32}. In the right panel (b) cyan, orange, red and blue colour exhibit points for which values of $m_{\phi_2}$ equals to $1.6\,m_{\phi_1}$, $1.8\,m_{\phi_1}$, $2\, m_{\phi_1}$ and $2.2\,m_{\phi_1}$ respectively. }
\end{center}
\end{figure}

In this section we  highlight the phenomenon of resonant semi-annihilation which is possible within  the multipartite $\mathbb{Z}_3$ scenarios. Note that for the process $\phi_1 \phi_1 \rightarrow \phi_2 \rightarrow \phi_1 h$  in Fig.~\ref{sf:semiannz32}, has a resonance in the vicinity of  $m_{\phi_1}\sim m_{\phi_2}/2$, where the semi-annihilation cross-section shoots up,  reducing relic density. This is a novel feature of the non-minimal $\mathbb{Z}_3$ model and should be contrasted with the Higgs resonance as this does not put any restriction on the DM mass. One can tune the NLSP mass approximately to achieve this for any value of DM mass $m_{\phi_1}$. The couplings involved here are $\lambda_{1h},\lambda_{2h}, \lambda_{12h}, \mu_1, \mu_2,\mu_{12}$ out of which only $\lambda_{1h}$ contributes to direct search.  We demonstrate this with a limited scanning in the region  of interest in the  parameter space as specified below,
\begin{equation}
\lambda_{1h}=\lambda_{2h}=0.001, ~ 0.0001 \leq \lambda_{12h} \leq 0.01, ~\mu_{i} \leq 2 m_{\phi_i},~ \mu_{12} \leq 2 m_{\phi_1}.
\label{eq:ps-rsa}
\end{equation}
In Fig.~\ref{fig:m2omega}  we show the variation in relic density due to change in $m_{\phi_2}$ for a fixed $m_{\phi_1}$ ($m_{\phi_1}= 300$~GeV). The pronounced resonance effect is  evident  near $m_{\phi_2} \sim 2  m_{\phi_1}$  where the relic density drops sharply\footnote{The features observed in the region $m_{\phi_2} > 2 m_{\phi_1} $ arises due to convergence issues in obtaining the thermally averaged cross section in {\tt micrOmegas} and do not signify any underlying physics}. Resonant semi-annihilation effect is also shown in $m_{\phi_1}-\Omega h^2$ plane for different choices of $m_{\phi_2}$ in Fig.~\ref{fig:m1omega} which clearly shows that the resonant semi-annihilation can contribute significantly for a large range of DM masses. The  parameters for the scan are identical to Eq.~\ref{eq:ps-rsa} except now we  vary the NLSP mass in the range  $
1.5~m_{\phi_1} \leq  m_{\phi_2} \leq  2.5~m_{\phi_1}$ and  $m_{\phi_1}$ between $125-1000$ GeV.
In both cases (Fig.~\ref{fig:m2omega} and Fig.~\ref{fig:m1omega}) the  couplings are  deliberately chosen so that the relic density is only satisfied at resonance semi-annihilation. This implies that   the  Higgs portal couplings are small, easily evading direct search constraints.

%
%%%%%%%%%%%%%%%%%%%%%
\subsubsection{Numerical Scans and Analysis}
\label{sssec:DDz32}
%%%%%%%%%%%%%%%%%%%%%
\begin{figure}[t]
$$
 \includegraphics[scale=0.5]{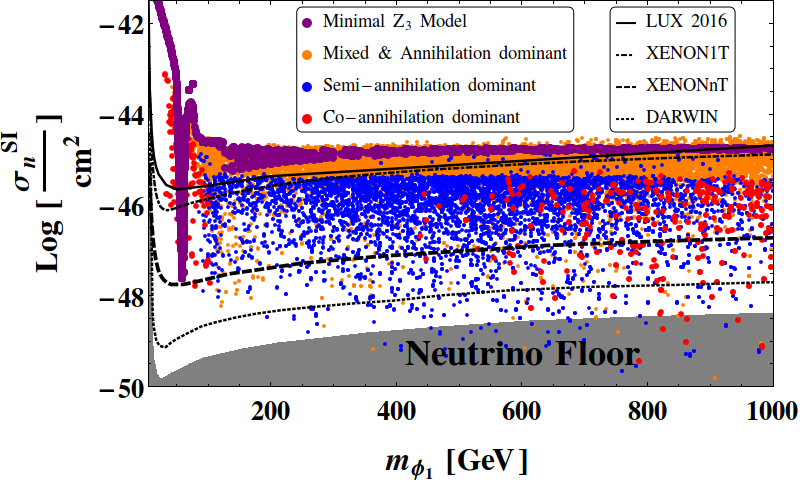}
$$
 \caption{Spin-independent DM-nucleon Direct Detection cross section for relic density allowed parameter space of two particles under $\mathbb{Z}_3$ model depicted {Eq.~\ref{eq:potz32}} as function of DM mass for parameters indicated in Eq.~\ref{eq:psz32}. LUX, XENON1T bound XENONnT and DARWIN sensitivities are indicated with Neutrino Floor  represented by gray shaded region \cite{Billard:2013qya}.} 
 \label{fig:DD-z3-2}
\end{figure}
To investigate the generic features of the non-minimal $\mathbb{Z}_3$ model we perform a large five parameter scan.  The relevant parameters are varied in the following range,
\begin{eqnarray}
10 ~{\rm GeV}< m_{\phi_1}< 1000~{\rm GeV},~2 \leq~\Delta{m}= m_{\phi_2}- m_{\phi_1} \leq 1000~ \rm GeV, \nonumber\\
   \mu_i \leq 2 m_i, ~\mu_{12}\leq  2 m_1 ,~
0.001 \leq \lambda_{1h}=\lambda_{2h} \leq 1 ,~0.1 ~\leq \lambda_{12h}\leq~ 1~.
\label{eq:psz32}
\end{eqnarray}

The tree level DM-nucleon coupling relevant for direct search experiments is still  mediated by the usual t-channel  exchange of  Higgs  and is proportional to the  coupling $\lambda_{1h}$ as in the minimal framework discussed in section~\ref{subsubsec:DDz22}.  We plot the relic density allowed parameter space points emerging from the scan in the spin independent direct search cross-section versus DM mass plane in Fig.~\ref{fig:DD-z3-2}.  We note  that there exist a lot of relic density allowed points beyond XENON1T \cite{Aprile:2017iyp} limit and proposed XENONnT \cite{Aprile:2015uzo} limit. They can go beyond DARWIN  sensitivity getting submerged into the neutrino floor for a wide range of DM mass. Once again, we  highlight the dominant underlying channels  by considering  three  regions:
\begin{itemize}
\item  Mixed  \& Annihilation dominant  (orange points),
\item Semi-annihilation dominant (blue points), 
\item Co-annihilation dominant (red points).
\end{itemize}
By dominant, we mean that more than 80$\%$ contributions to the required annihilation cross-section appears from these channels respectively. Parameter space for which both semi-annihilation and co-annihilation is sub dominant represented by orange color and labeled as Mixed \& Annihilation dominant. 

\begin{figure}[t]
\begin{center}
\subfloat[\label{sf:m1m2rl}]{
\includegraphics[scale=0.26]{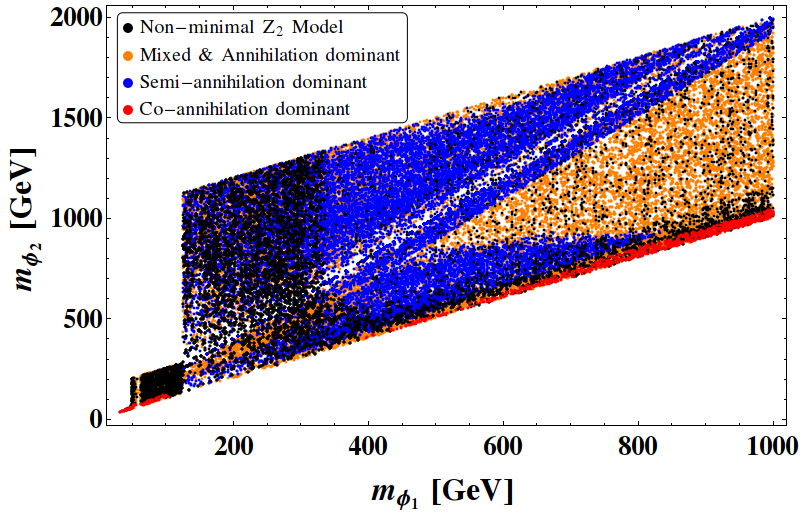}}
\subfloat[\label{sf:m1m2withlux}]{
\includegraphics[scale=0.26]{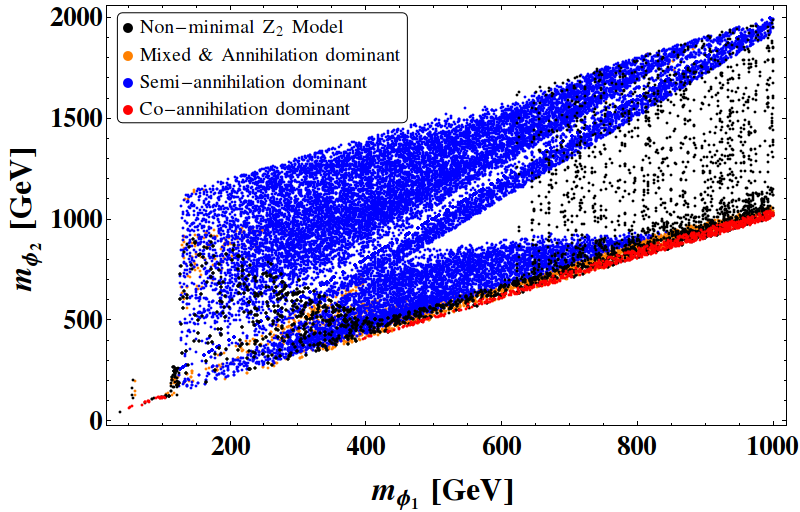}}\\
\subfloat[\label{sf:m1m2withxenon1t}]{
\includegraphics[scale=0.26]{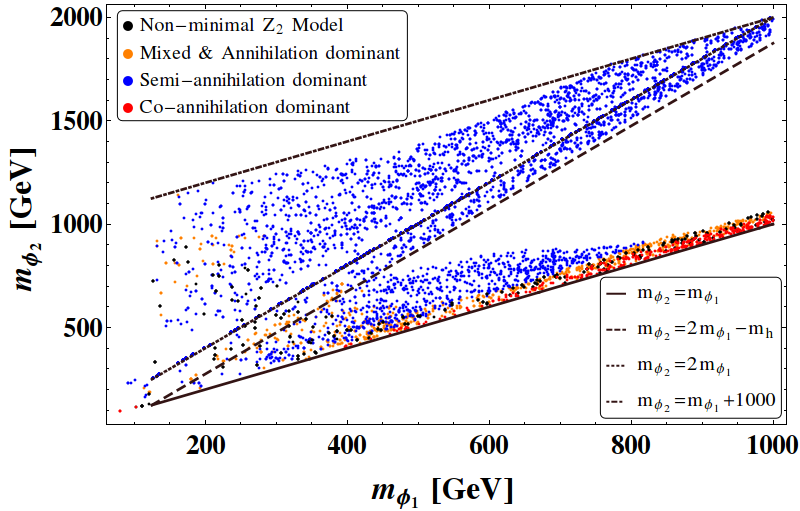}}~
\caption{(a) Relic density(PLANCK) allowed points, (b) Relic density and XENON1T allowed points, (c) Relic density and XENONnT allowed points in $m_{\phi_1}-m_{\phi_2}$ plane for the parameters prescribed in Eq. \ref{eq:psz32}, for two particles under $\mathbb{Z}_3$ model depicted in {Eq.~\ref{eq:potz32}}. The kinematic boundary lines are shown in the bottom panel.} 
\label{fig:m1m2}
\end{center}
\end{figure}
To demonstrate the importance of  co-annihilation and resonant semi-annihilation  in simultaneously addressing the relic density prediction while avoiding stringent direct detection constraints  we plot the allowed points on the $m_{\phi_1}-m_{\phi_2}$  in Fig. \ref{fig:m1m2}. On the top left panel Fig.~\ref{sf:m1m2rl}, we show only relic density allowed parameter space. Those satisfying both relic density constraint and XENON1T \cite{Aprile:2017iyp} limit are shown in Fig.~\ref{sf:m1m2withlux}. In the bottom panel \ref{sf:m1m2withxenon1t}, we depict the parameter space that will survive even XENONnT \cite{Aprile:2015uzo} limit.  We again use different colors to identify points with the dominant underlying process that contribute to relic density calculations. The plots clearly shows that there are two distinctive domains in the parameter space where this model is expected to remain relatively unconstrained by the present and proposed direct detection experiments. One of them is bunched in the region $ m_{\phi_2} \sim m_{\phi_1}$ and is dominated by the co-annihilation of DM with the NLSP. This also represent the only region where we can expect the multipartite $\mathbb{Z}_2$ models to survive. An entire new wing is obtained around $ m_{\phi_2} \approx 2 m_{\phi_1}$, for the resonant semi-annihilation  channel. These kinematic boundary lines like $ m_{\phi_2} = m_{\phi_1}$ and $ m_{\phi_2} = 2 m_{\phi_1}$ shown in the bottom panel of Fig. \ref{fig:m1m2}. This is an entire new region of unconstrained parameter space that is obtained in the non-minimal $\mathbb{Z}_3$ model. The two regions while remaining unconstrained by direct searches should lead to distinct different consequences in the indirect searches  \cite{Arcadi:2017vis}. The resonant semi-annihilation region can lead to striking indirect detection signals while the co-annihilation processes do not contribute to the this signal strength at all. A detailed study of this will be carried out elsewhere.

 \begin{figure}[t]
$$
 \includegraphics[scale=0.36]{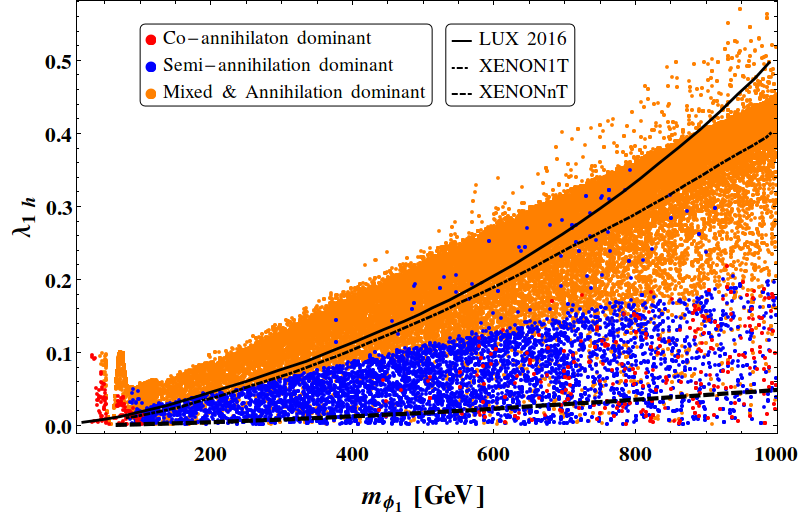}
$$
 \caption{Relic density allowed parameter space in $m_{\phi_1}-\lambda_{1h}$ plane for two particles under $\mathbb{Z}_3$ model depicted in {Eq.~\ref{eq:potz32}}. Semi-annihilation (blue), Co-annihilation (red) and Mixed \& annihilation (orange) dominated regions are indicated separately. } 
 \label{fig:lh-m1-z3-2}
\end{figure}

The relic density allowed points projected in the plane of $m_{\phi_1}-\lambda_{1h}$  is shown in Fig.~\ref{fig:lh-m1-z3-2}. One can clearly see, that  most of mixed \& annihilation dominated region of non-minimal $\mathbb{Z}_3$ is disfavored by current XENON1T limits. For large  contribution from co-annihilations and semi-annihilation (red and blue points which superpose on each other), the required $\lambda_{1h}$ coupling can be brought down significantly and all these regions will be allowed by present direct search constraints. 

From the experience with the $\mathbb{Z}_2$ case we can conclude that inclusion of more states in the dark sector charged under $\mathbb{Z}_3,$ will add  copies of the channels that are operative in this model. Proliferation of  annihilation modes in the $N$-state $\mathbb{Z}_3$ model  will contribute to further relaxation in the direct detection constraints for the relic density allowed region. It is expected that similar reduction of the direct detection cross section can now be achieved for  smaller values of the co-annihilation and semi-annihilation couplings as compared  to the two scalar  case   that has been numerically explored in this section.

%%%%%%%%%%%%%%%%%%%%%%%%%%%%%%%%
\subsection{Two Component DM in $\mathbb{Z}_3\times \mathbb{Z}_3^{'}$ :  DM Exchange}
\label{sec:teo-comp-z3}
%%%%%%%%%%%%%%%%%%%%%%%%%%%%%%%%%
For completion we consider the  two component DM that transform under two different $\mathbb{Z}_3$. The $\mathbb{Z}_3\times  \mathbb{Z}_3^{'}$ invariant Lagrangian is  given by,
\bea\label{eqn:potz3z3p}
-{\mathcal{L}}_{DM-Higgs} &=&-\mu_{H}^2 (H^{\dagger}H-\frac{v^2}{2} )+{\ \lambda_H}(H^{\dagger}H-\frac{v^2}{2} )^2+ m^{2}_{\phi_{1}} \phi^{*}_{1}\phi _{1}+m^{2}_{\phi _{2}} \phi^{*}_{2}\phi _{2} \nonumber \\
 &&+\mu_{1} (\phi^{3}_{1}+\mbox{h.c})+\mu_{2}( \phi^{3}_{2}+\mbox{h.c} )+\lambda_{1s}(\phi^{*}_{1}\phi _{1})^{2}+\lambda_{2s}(\phi^{*}_{2}\phi _{2})^{2}+\lambda_{e}(\phi^{*}_{1}\phi _{1}) (\phi^{*}_{2}\phi _{2})\nonumber \\
 && +\lambda_{1h}(\phi^{*}_{1}\phi _{1})(H^{\dagger}H-\frac{v^2}{2}) + \lambda_{2h}(\phi^{*}_{2}\phi _{2}) (H^{\dagger}H-\frac{v^2}{2}). 
\eea
 Unlike in the previous section, here we obtain a two component DM. On top of  number changing processes  like  annihilations and semi-annihilations, there will be exchange  processes  as shown in the Feynman graphs in Fig.~\ref{fig:Z32_dig}. Note that co-annihilation and resonant semi-annihilation process are not present in this minimal setup. The existence of two DM candidates requires large reduction of number densities and this is more disfavored by direct search experiments. To maximize the impact of semi-annihilation we use the indulgent limit of  $\mu_i \lesssim7.4 m_{\phi_i}$ \cite{Belanger:2012zr}.
 
\begin{figure}[t]
%\begin{itemize}
 \begin{center}\subfloat[\label{sf:semiannZ32}]{
     \begin{tikzpicture}[line width=0.5 pt, scale=1.]
       %s channel  
        \draw[dashed] (-1,1)--(0,0);
	\draw[dashed] (-1,-1)--(0,0);
	\draw[dashed] (0,0)--(1.5,0);
	\draw[dashed] (1.5,0)--(2.5,1);
	\draw[dashed] (1.5,0)--(2.5,-1);
	\node  at (-1.4,-1) {$\phi_{i}$};
	\node at (-1.4,1) {$\phi_{i}$};
	\node [above] at (0.75,0) {$\phi_{i}$};
	\node at (2.9,1) {$h$};
	\node at (2.9,-1) {$\phi^{*}_{i}$};
      %t channel
      \draw[dashed] (5,1)--(6,0);
	\draw[dashed] (6,0)--(6,-1.0);
	\draw[dashed] (5,-2.0)--(6,-1.0);
	\draw[dashed] (6,0)--(7,1);
	\draw[dashed] (6,-1.0)--(7,-2.0);
	\node  at (4.6,1) {$\phi_{i}$};
	\node at (7.4,1) {$\phi_{i}$};
	\node [right] at (6,-0.5) {$\phi_{i}$};
	\node at (7.2,-2) {$h$};
	\node at (4.6,-2) {$\phi^{*}_{i}$};
	%\label{fd:semi}
     \end{tikzpicture}}
 \end{center}
 
 \begin{center}\subfloat[\label{sf:exchangeZ32}]{
    \begin{tikzpicture}[line width=0.5 pt, scale=1.]
    %For phi,phi -> h,h point
        \draw[dashed] (-5,1)--(-4,0);
	\draw[dashed] (-5,-1)--(-4,0);
	\draw[dashed] (-4,0)--(-3,1);
	\draw[dashed] (-4,0)--(-3,-1);
	\node at (-5.4,1) {$\phi_{1}$};
	\node at (-5.4,-1) {$\phi^{*}_{1}$};
	\node at (-2.8,1) {$\phi_{2}$};
	\node at (-2.8,-1) {$\phi^{*}_{2}$};
	%For phi,phi ->h,h
        
        \draw[dashed] (-1,1)--(0,0);
	\draw[dashed] (-1,-1)--(0,0);
	\draw[dashed] (0,0)--(1.5,0);
	\draw[dashed] (1.5,0)--(2.5,1);
	\draw[dashed] (1.5,0)--(2.5,-1);
	\node  at (-1.2,-1) {$\phi^{*}_{1}$};
	\node at (-1.2,1) {$\phi_{1}$};
	\node [above] at (0.75,0) {$h$};
	\node at (2.7,1) {$\phi_{2}$};
	\node at (2.7,-1) {$\phi^{*}_{2}$};
   \end{tikzpicture}}
 \end{center}
 \caption{Feynman diagrams  : a) Semi-annihilation ($i=1,2$) and b) DM DM exchange.}
 \label{fig:Z32_dig}
 \end{figure}
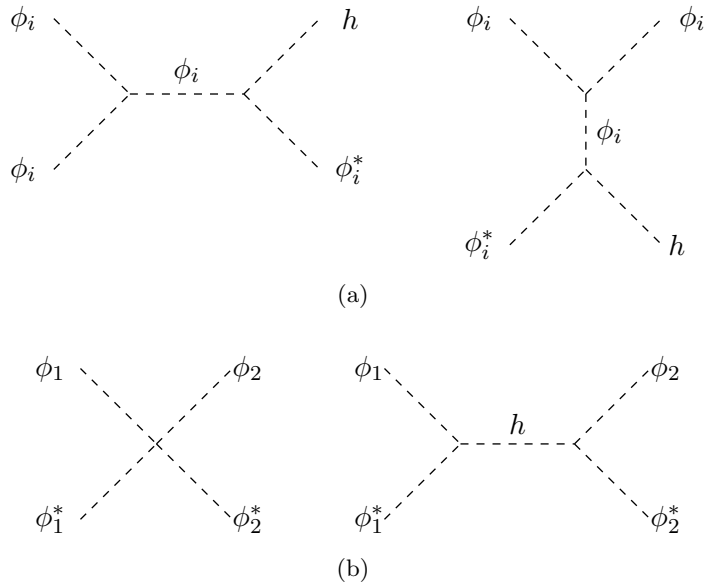

The coupled Boltzmann equation for this two component DM case can be written as \cite{Aoki:2012ub},
\bea
\frac{dn_{\phi_1}}{dt}+3Hn_{\phi_1}&=&-\langle \sigma v \rangle_{\phi_1\phi_1\to SM}(n_{\phi_1}^2  
-{n_{\phi_1}^{eq}}^2)-\frac{1}{2}\langle \sigma v \rangle_{\phi_1\phi_1\to\phi_1 SM}(n_{\phi_1}^2-n_{\phi_1}n_{\phi_1}^{eq})  \nonumber \\
&&- \langle \sigma v \rangle_{\phi_1\phi_1\to \phi_2 \phi_2}[n_{\phi_1}^2-(\frac{n_{\phi_1}^{eq}}{n_{\phi_2}^{eq}})^2n_{\phi_2}^2] +\langle \sigma v \rangle_{\phi_2\phi_2\to \phi_1\phi_1}[n_{\phi_2}^2-(\frac{n_{\phi_2}^{eq}}{n_{\phi_1}^{eq}})^2n_{\phi_1}^2]~, \nonumber \\
\frac{dn_{\phi_2}}{dt}+3Hn_{\phi_2}&=&-\langle \sigma v \rangle_{\phi_2\phi_2\to SM}(n_{\phi_2}^2-{n_{\phi_2}^{eq}}^2)
-\frac{1}{2}\langle \sigma v \rangle_{\phi_2\phi_2\to\phi_2 SM}(n_{\phi_2}^2-n_{\phi_2}n_{\phi_2}^{eq})\\
&&+ \langle \sigma v \rangle_{\phi_1\phi_1\to \phi_2 \phi_2}[n_{\phi_1}^2-(\frac{n_{\phi_1}^{eq}}{n_{\phi_2}^{eq}})^2n_{\phi_2}^2] -\langle \sigma v \rangle_{\phi_2\phi_2\to \phi_1\phi_1}[n_{\phi_2}^2-(\frac{n_{\phi_2}^{eq}}{n_{\phi_1}^{eq}})^2n_{\phi_1}^2]\nonumber.
\eea
We have inserted this two component framework into {\tt{micrOmegas}} and looked for relic density allowed parameter space using the algorithm given in Appendix.~\ref{apnd:mthd}. However,  the direct detection cross section has been  calculated manually  using following formula \cite{Bhattacharya:2013hva,Cao:2007fy},
\begin{eqnarray}
 \sigma_{eff}^{i}= (\frac{\Omega_i}{\Omega_T})(\sigma_{n}^{SI})_i=\frac{\Omega_i}{\Omega_T}\frac{\lambda_{ih}^2 f_n^2}{4\pi } \frac{\mu_n^2 m_n^2}{m_h^4m_{\phi_i}^2}~~~~~~~~(i=1,2),
 \label{eq:dd1}
 \end{eqnarray} 
where the DM-nucleon cross section in detailed in Section~\ref{subsubsec:DDz22}.

 The key  features that emerges out of the analysis is shown in Fig.~\ref{fig:DD1-z3-z3p}  obtained by scanning over the following parameters,
 \bea
 \label{z2z2para}
125 \leq m_1 \leq 500  , ~500 \leq m_2 \leq 1000  , ~5m_i \leq \mu_1 =\mu_2 \leq 7.4 m_i , \nonumber \\
 0.01 \leq \lambda_{1h}=\lambda_{2h} \leq 0.1, ~0.5 \leq \lambda_{e}\leq 1.5
\eea
 In the allowed parameter space we have  scenarios where the two components are separated in mass by $>300$ GeV and remains in the range for the next generation XENON experiments. This provides the tantalizing possibility of detecting two different DM particles in direct detection experiment signaling a multi-component dark sector. This should be contrasted with the  $\mathbb{Z}_2 \times  \mathbb{Z}_2'$ model detailed in \cite{Bhattacharya:2016ysw}, where the  allowed DM  pushes both components to be beyond $\gtrsim 400$ GeV. This forbids the concurrent discovery of two DM candidates in near future. 
 
 A two component DM model with $\mathbb{Z}_3 \times  \mathbb{Z}_3'$ can further be extended with more dark sector particles transforming either $\mathbb{Z}_3$ or $\mathbb{Z}_3'$ or both to have co-annihilation and resonant semi annihilation to evade direct search constraints to a great extent. However the phenomenology will be simply guided by the analogy in section \ref{subsec:semi-and-co} and \ref{sec:teo-comp-z3} together and not illustrated here.

\begin{figure}[t]
 $$
 \includegraphics[height=4cm]{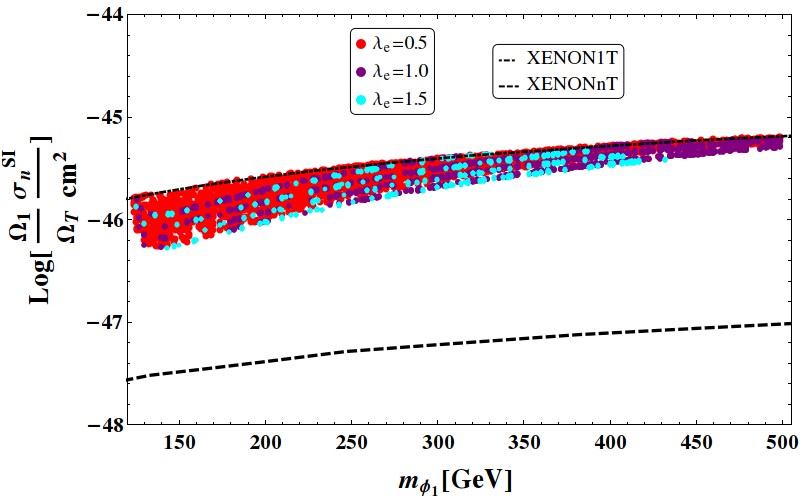}~~~
 \includegraphics[height=4cm]{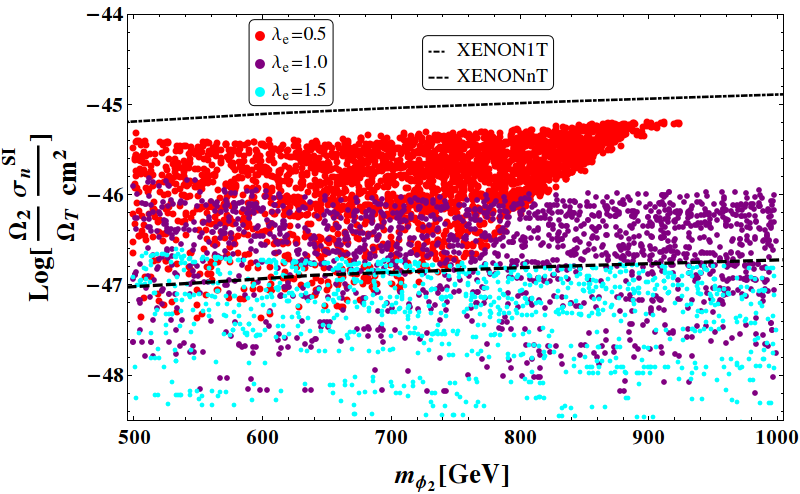} 
 $$
  \caption{Spin-independent DM-nucleon effective cross-section vs DM mass for relic density allowed points in $\mathbb{Z}_3\times \mathbb{Z}_3^{'}$ model indicated in Eq. \ref{eqn:potz3z3p} where both $\phi_1$ and $\phi_2$ are allowed by XENON1T limit for the scan over parameters specified in Eq.~\ref{z2z2para}.} 
  \label{fig:DD1-z3-z3p}
 \end{figure} 
 %%%%%%%%%%%%%%%%%%%%%%%%%%%%%%%%%%%
 \section{Brief Sketch of Vaccum Stability and Unitarity Constraints}
 \label{nnnn}
 
Here we briefly summerize the  constraints on the parameter space of the models considered above from tree level unitarity and vacuum stability. We will consider the $\mathbb{Z}_2$ and $\mathbb{Z}_3$ models in turn.

The stability of the minimal $\mathbb{Z}_2$ Higgs portal scalar DM models have been studied in \cite{Lerner:2009xg}, while  the extension to $\mathbb{Z}_2 \times \mathbb{Z}^{'}_2$ is straight-forward. For the multipartite $\mathbb{Z}_2$ stabilized Higgs portal models as described in  the Lagrangian given in Eq.~\ref{potz22} the  condition for global minimum can be obtained following \cite{Kannike:2016fmd}. The potential given in Eq. \ref{potz22} can be recast in the following form
\begin{equation}
\label{eq:potz22-re}
V(\phi_1,\,\phi_2,\, H) =\lambda_H \lvert H \rvert^4+ \mathsf{M}(\phi_1,\,\phi_2) \lvert H \rvert^2+v(\phi_1,\,\phi_2),
\end{equation}
where,  $$\mathsf{M}(\phi_1,\,\phi_2)= \frac{1}{2}\lambda_{1h}\phi_1^2+\frac{1}{2}\lambda_{2h}\phi_2^2 +\lambda_{12h}\phi_1\phi_2 $$  and
$$v(\phi_1,\,\phi_2)=\frac{\lambda_{e_1}}{4}\phi_1^2\phi_2^2+\frac{\lambda_{e_2}}{3!}\phi_1^3\phi_2+\frac{\lambda_{e_3}}{3!}\phi_1\phi_2^3 + \frac{\lambda_{1s}}{4!}\phi_1^4+\frac{\lambda_{2s}}{4!}\phi_2^4$$
Since the fields have positive nonzero mass we only consider the  quartic terms in fields. Condition for the stability of the potential in Eq. \ref{eq:potz22-re}  implies that the discriminant of it with respect to $\lvert H \rvert^2$ is positive. This gives rise to the following conditions,
\begin{eqnarray}
\lambda_{1s}> 0,~~ \, 
\lambda_{2s}> 0,~~ \, &\nonumber \\
6\l_{12h}\l_{1h}(\l_{e_2}+\l_{e_3})+3\l_{e_1}\l_{H}\l_{s}-3\l_{12h}^2\l_{s}
&-2\l_{H}\left(\l_{e_2}^2+\l_{e_3}^2\right)>0 
\end{eqnarray}
where, for simplicity,  we have assumed $\lambda_{1h}=\lambda_{2h}$, $\lambda_{1s}=\lambda_{2s}=\lambda_s \gg \lambda_{1h}$. These are the condition for which  the potential given in Eq. \ref{eq:potz22-re} has only one global minimum. As explained in main text $\lambda_{e_i}, \lambda_s$ does not play any role in DM phenomenology, so by tuning them to be sufficiently large, but well within the perturbative limits,  one can easily ensure the stability of the potential in all regions of the parameter space that has been scanned.
\begin{table}[t]
\centering
\resizebox{\textwidth}{!}{%
\begin{tabular}{|c|c|c|}
\hline
Model                    & Lagrangian & Tree level unitarity constraints                                                                                                                                                                                                                                                                                                                             \\ \hline
Two particle under $\mathbb{Z}_2$                   & Eq. \ref{potz22}       & \begin{tabular}[c]{@{}c@{}}$\l_H < 4 \pi, \,\left( \l_{1h}+\l_{2h}\pm\sqrt{4 \l^2_{12h}+\left(\l_{1h}-\l_{2h}\right)^2}\right) < 16 \pi,$ \\ $x< 8\pi$ where $x$ is the solution of the equation \ref{eq:unitarity-z2-2}.
\end{tabular}                                                                                                                                             \\ \hline
One particle under $\mathbb{Z}_3$ & Eq. \ref{eq:mz3}       & \begin{tabular}[c]{@{}c@{}}\newline $\lambda_{H}< 4 \pi,\,\lambda_{1h}< 8 \pi,\, \lambda_{1s}< 4 \pi, $\\ $2\lambda_{1s}+3\lambda_{H}\pm\sqrt{2\lambda_{1h}^2+(2\lambda_{1s}-3\lambda_{H})^2}<8 \pi$\end{tabular}                                                                                                                                         \\ \hline
Two particle under $\mathbb{Z}_3$ & Eq.  \ref{eq:potz32}      & \begin{tabular}[c]{@{}c@{}}$\lambda_{H}<4 \pi,\,\left(\lambda_{1h}+\lambda_{2h}\pm\sqrt{4\lambda_{12h}^2+(\lambda_{1h}-\lambda_{2h})^2}\right)<16 \pi,$\\ $\,\lambda_{e}<\frac{8 \pi}{3},\,\left(\lambda_{1s}+\lambda_{2s}\pm\sqrt{(\lambda_{1s}-\lambda_{2s})^2+4\lambda^2_{e}}\right)<8 \pi,$\\ $y<8 \pi$  where $y$ is the solution of the equation \ref{eq:unitarity-z3-2}.
\end{tabular} \\ \hline
$\mathbb{Z}_3 \times \mathbb{Z}^{'}_3$                    & Eq. \ref{eqn:potz3z3p}       & \begin{tabular}[c]{@{}c@{}}$\lambda_{H}<4 \pi,\, \lambda_{1h}<8 \pi,\, \lambda_{2h}< 8 \pi,\, \lambda_{1s}< 4\pi,\, \lambda_ {2s}< 4\pi, \lambda_ {e}< 8\pi ,$\\ $ z<8\pi $ where $z$ is the solution of the equation \ref{eq:unitarity-z3-z3p}.
\end{tabular}                                                                                                                                          \\ \hline
\end{tabular}%
}
\caption{Tree level unitarity constraints for discussed models.}
\label{table:unitarity}
\end{table}

For tree level constraint from unitarity we utilize  the  Lee, Quigg and Thacker (LQT) method \cite{Lee:1977eg} widely used in various BSM context  \cite{Horejsi:2005da, Das:2014fea, Mondal:2015fja, Chakrabarty:2014aya, Bhattacharyya:2015nca}. The scattering processes which goes through dimensionful couplings have propagator suppression remaining relatively unconstrained. The constrains on the dimension less parameters are obtained as detailed in Appendix \ref{unitcons} and is given in  table \ref{table:unitarity}. 
It is easily seen that for region of  parameter space scanned the constrained are satisfied.  For example  in a typical  choice of parameters in the region of interest:
$m_{\phi_1}=775~\mbox{GeV}, ~ m_{\phi_2}=778 ~\mbox{GeV}, ~  \l_{1h}=\l_{2h}=0.002, ~ \l_{12h}=0.35,~ \l_{e_i}=1 ~\mbox{and}  ~\l_{is}=1$ the absolute values of the solutions of Eq.~\ref{eq:unitarity-z2-2} are $ \sim 0, ~0.1,~ 1.2~\mbox{and} ~2.$

Vaccum stability for the  potential having $\mathbb{Z}_3$ symmetry has been studied extensively in  \cite{Belanger:2012zr}. In this case we can have several possible stationary points. The stationary points are 
$(i)~\left\langle H \right\rangle=\left\langle \phi \right\rangle=0$,
$(ii)~\left\langle H \right\rangle \neq 0 $ and $\left\langle \phi \right\rangle=0$,
$(iii)~\left\langle H \right\rangle = 0 $ and $\left\langle \phi \right\rangle \neq 0$,
$(iv)~\left\langle H \right\rangle \neq 0 $ and $\left\langle \phi \right\rangle \neq 0$. The desired SM vaccum with a stable DM is defined by $(i)$. Assuming all the quartic couplings are positive in the Lagrangian given by \ref{eq:pot} condition for this to be the global minima sets an upper bound on the trilinear coupling $\mu_1 \lesssim 2 m_{\phi_1}$. A metastable vaccum with the desired property and lifetime greater than the age of the universe,  relaxes the bound to $\mu_1 \lesssim 7.4 m_{\phi_1}$. In the $\mathbb{Z}_3 \times \mathbb{Z}^{'}_{3}$ model discussed in section \ref{sec:teo-comp-z3} these limits gets extended on every trilinear coupling $\mu_i$ in the Lagrangian defined in Eq. \ref{eqn:potz3z3p}. The situation for multipartite $\mathbb{Z}_3$ model is more involved due to the additional trilinear couplings involving multiple dark sector state. A detailed study entail the estimation of lifetime of the desired metastable vacuum which essentially constraints the trilinear couplings to be smaller  and  relatively large quartics. Therefore we adhere to a conservative limit of $\mu \lesssim 2 m_\phi.$

Following the algorithm portrayed in Appendix \ref{unitcons} unitarity constraint on the parameter space for the Lagrangian stated in Eq. \ref{eq:potz32} are given in table \ref{table:unitarity}. It is clear that the given constraints are satisfied for the region scanned. For  a representative  parameter point with $\l_{1h}=\l_{2h}=0.001, ~ \l_{12h}=0.5,~ \l_{e}=1 ~\mbox{and}  ~\l_{is}=1,$   the solutions of Eq. \ref{eq:unitarity-z3-2} are $1,\,3,\, 5,\,5.2$. An identical discussion also ensure unitary behavior of the  $\mathbb{Z}_3 \times \mathbb{Z}^{'}_3$ model for the constraints  given in table \ref{table:unitarity}.

%%%%%%%%%%%%%%%%%%%%%%%%%%%%%%%%%%%%%%%%%%%%
\section{Conclusions}
\label{sec:summary}
%%%%%%%%%%%%%%%%%%%%%%%%%%%%%%%%%%%%%%%%%%%%
Ever increasing precision of direct detection experiments set  considerable constraints  on the otherwise well motivated scalar Higgs portal DM framework.
A significant region of the parameter space  within the  minimal   model has been ruled out by the current bounds on DM-nucleon cross section from  LUX 2016 and XENON1T. Except the tuned Higgs  pole region, this model is on the verge of being precluded by the next generation experiments.  In this paper we perform a systematic study of simple extensions of the minimal Higgs portal framework in terms of their viability of surviving next generation direct detection experiments while saturating the DM relic abundance estimates. Here we have considered both enlarging the stabilizing symmetry from $\mathbb{Z}_2 \rightarrow \mathbb{Z}_3$ and introducing multipartite dark sectors.

In  the two particle  $\mathbb{Z}_2$ framework we can have novel DM number changing processes like,   co-annihilation and mediated annihilation. Phase space barrier of co-annihilation and final state Higgs in the mediated annihilation  channels  make these topologies insensitive to direct search experiments  though they contribute to freeze out.  This provides  a handle to disentangle   these two phenomenon which are usually correlated within the WIMP paradigm. A  release  of  imminent tension with direct detection experiments  is obtained, allowing them to survive upto and even beyond the  proposed sensitivity of  DARWIN. We perform an extensive numerical simulation of the model to study the interplay of these processes in alleviating the direct detection bounds. We observe that mediated annihilation  processes are  effective for the DM mass below $400$ GeV, while co-annihilation plays  a similar role for larger masses,  provided that the DM and the NLSP are relatively degenerate.  Expectedly increasing the number of states contributing to these processes by populating the dark sector with more species  will further alleviate the tension.

Within the  minimal $\mathbb{Z}_3$  scalar singlet Higgs portal models  the semi-annihilation  processes are brought into play.  This can help sustain the model beyond XENON1T  for DM mass above the  Higgs mass and below $\sim 400$ GeV. However, the semi-annihilation processes are constrained from above by vacuum stability considerations and will be unable to survive an absence  at XENONnT. Whereas a non-minimal $\mathbb{Z}_3$ facilitate the possibility of co-annihilation, mediated annihilation,  resonant semi-annihilation etc.   which are efficient in allowing the DM direct detection cross section to as low as  the sensitivity of  direct detection experiments set by the ambient neutrino flux at these  class of experiments. We perform a multidimensional numerical scan to present the allowed parameter space  and highlight the role of various underlying processes that contribute to the relic density and  direct detection calculations. We briefly discuss the possible constraints on the allowed parameter space from tree level unitarity and vacuum stability conditions.

We conclude that a no show at next generation experiments will push the  intermediate scale Higgs portal DM paradigm into the multipartite era. Having multi-component DM stabilized with individual symmetries admittedly  worsen the situation.  However, the prognosis is much better for models where there is a multipartite dark sector all charged under the same stabilizing symmetry, where the lightest charged state is the cosmologically stable  DM candidate. In this case we find that there are two interesting possibilities  of co-annihilation and resonant semi-annihilation, that enable effective reduction of the DM-nucleon cross section to remain virtually unconstrained by direct detection while driving the relic density to the right ball park.  Co-annihilation require a relative degeneracy in the DM and the NLSP  while the possibility of the resonant semi-annihilation is effective provided we admit a  tuning of the form $m_{NLSP} \sim 2 m_{DM}.$ Interestingly these two possibilities while being insensitive to probing through direct detection experiments will provide strikingly different signatures for indirect detection experiments.

%%%%%%%%%%%%%%%%%%%%%%%%%%%%%%%%%%%%%%%
\paragraph*{Acknowledgments\,:} 
%%%%%%%%%%%%%%%%%%%%%%%%%%%%%%%%%%%%%%%
We thank Ujjal Kumar Dey and  Sayan Das Gupta for discussions.  SB would like to acknowledge the DST-INSPIRE research grant IFA13-PH-57 at IIT Guwahati. SB and PG acknowledges the hospitality at IIT Kharagpur where the work was planned. PG and TNM would  like to thank MHRD, Government of India for  research fellowship.  TSR acknowledges
the hospitality provided by ICTP, Italy, under the Associate
program, during the completion of the project. TSR is partially supported by the Department of Science
and Technology, Government of India, under the Grant Agreement number IFA13-PH-74
(INSPIRE Faculty Award).
%%%%%%%%%%%%%%%%%%%%%%%%%%%%%%%%%%%%%%%%%%%%%%%%%%%
\appendix
%%%%%%%%%%%%%%%%%%%%%%%%%%%%%%%%%%%%%%%%%%%%%%%%%%%

%%%%%%%%%%%%%%%%%%%%%%%%%%%%%%%%%%%%%%%%%%%%%%%%%%%%%%%%%
\section{Numerical Procedure}
\label{apnd:mthd}
In this appendix we briefly summarize  the numerical method  we have followed to scan the parameter space through out this paper. Each model  has been implemented utilizing the  {\tt LanHEP} \cite{Semenov:2014rea} platform. The extract model files from {\tt LanHEP} have been sanity checked using   {\tt CalCHEP} \cite{Belyaev:2012qa}. Then we implement model files into {\tt micrOmegas} \cite{Belanger:2014vza} which have been used for making extensive scan of the models presented in the text.  {\tt micrOmegas} has been used to  calculate  both the relic density $(\Omega h^2)$ and spin independent direct detection cross section $(\sigma_n^{\rm SI})$. We  also extract the information for the  contribution  from different channel to relic density for every parameter point simulated. In the case of the   two component DM presented in section \ref {sec:teo-comp-z3},  the    direct detection cross section is calculated
utilizing the analytical expressions given in Eq.~\ref{eq:dd1},  while {\tt micrOmegas}  has been used to obtain  relic density values.

\section{NLSP decay}
\label{apnd:decay}
\begin{figure}[htb!]
\label{decay_diag1}
\begin{center}
    \begin{tikzpicture}[line width=0.5 pt, scale=1.5]
	%For phi,phi ->f,f
        \draw[dashed] (-5,1)--(-4,1);
	\draw[dashed] (-4,1)--(-2.6,0.2);
	\draw[dashed] (-4,1)--(-3,1.5);
	\draw[solid] (-3,1.5)--(-2.5,2.0);
	\draw[solid] (-3,1.5)--(-2.5,1.2);
	\node  at (-5.2,1) {$\phi_2$};
	\node at (-2.4,0.2) {$\phi_1$};
	\node [above] at (-3.5,1.3) {$h$};
	\node at (-2.3,2.0) {$\rm SM$};
	\node at (-2.3,1.2) {$\rm SM$};
     \end{tikzpicture}
 \end{center}
 \caption{ $\phi_2$ decay to $\phi_1$ in presence of $\lambda_{12h}$. }
 \end{figure}
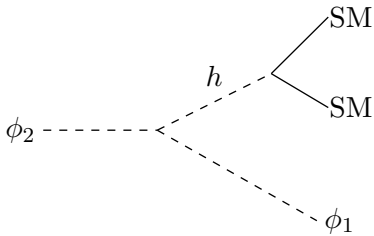

Here we point out lifetime of Next to Lightest Stable Particles which undergoes a three body decay.
The decay width for the heavier particle ($\phi_2$) to DM ($\phi_1$) is given by 
 \begin{eqnarray}
  \Gamma (\phi_2 \to \phi_1 \,X ) =\frac{1}{16 m_{\phi_2}(2 \pi)^3}\frac{\lambda_{12h}^2 m_f^2[(m_{\phi_2}-m_{\phi_1})^2-4 m_{f}^2]}{m_h^4}\Big[\frac{m_{\phi_1}^2(m_{\phi_2}-m_{\phi_1})}{m_{\phi_1}}\nonumber \\
  +\frac{(m_{\phi_2}-m_{\phi_1})^2}{2}-m_{\phi_1}^2 \ln(\frac{m_{\phi_2}}{m_{\phi_1}})\Big].
 \end{eqnarray}

 And decay time,$\tau(\phi_2 \to \phi_1 \,X )=1/\Gamma(\phi_2 \to \phi_1 \,X)$. We would like to evaluate the limit on the parameters of the model so that the decay width is not larger than the age of the universe ($0.66 \times 10^{42}$ $~{\rm GeV}^{-1}~$). A simple estimation shows  that within DM mass  $m_{\phi_1}$, $10-10^4$ GeV, the  limit on the  coupling, $\lambda_{12h}\sim 10^{-13}$ . In our analysis, we have consistently used the co-annihilation coupling  $\lambda_{12h} \sim 0.1 -1.0$. For this values the  heavier component $\phi_2$  decays promptly to $\phi_1$.

 \section{Unitarity Constraints} \label{unitcons}
The amplitude of a scattering  processes can be written in terms of Legendre polynomial as 
$$\mathcal{M}(\theta)=16 \pi \sum^{l=\infty}_{l=0}a_l (2l+1) \mathit{P}_l (\cos \theta)$$
In high energy limit the $a_0$ partial wave will  determine the leading energy dependence of  the  scattering processes. Therefore the unitarity constraint  on the process translates to $\lvert\rm Re~a_0 \rvert<1/2$. This imply the following bound on scattering amplitude \cite{Horejsi:2005da}
\begin{equation}
\label{eq:unitarirty}
\lvert \mathcal{M} \rvert < 8 \pi
\end{equation}
In our analysis we will confine ourself to two particle scattering processes. So the assignment is to calculate amplitude of all possible $2 \to 2$ scattering processes and employ Eq. \ref{eq:unitarirty} on the eigenvalues of amplitude matrix. \\
\textbf{Two particle under $\mathbb{Z}_2$:}\\
The model defined by the Lagrangian given in Eq.~\ref{potz22} have $11$ neutral two particle states i.e.
$$w^+w^-,~ \frac{hh}{\sqrt{2}},~  \frac{zz}{\sqrt{2}}, \frac{\phi_1 \phi_1}{\sqrt{2}},~ \frac{\phi_2 \phi_2}{\sqrt{2}},~ \phi_1 \phi_2,~ h z,~ h\phi_1,~  h\phi_2,~ z\phi_1,~ z\phi_2$$
 and $4$ singly charged particle states i.e.
$$ w^+h,~  w^+z,~ w^+\phi_1,~ w^+\phi_2.$$ Therefore for neutral states  the $11 \times 11$ amplitude matrix is given by
\begin{equation}
\mathcal{M}_{\rm NC}= \begin{pmatrix}
\mathcal{A}_{7\times 7} & 0 & 0 \\ 0 & \mathcal{B}_{2\times 2} & 0 \\ 0 & 0 & \mathcal{B}_{2\times 2}
\end{pmatrix} \,,
\label{eq:matrix-nc}
\end{equation} Where 
\begin{eqnarray*}
\mathcal{A} &=&
  \bordermatrix{
  &\m w^+ w^-&\m  \frac{hh}{\sqrt{2}}&\m \frac{zz}{\sqrt{2}}&\m \frac{\phi_1 \phi_1}{\sqrt{2}}&\m \frac{\phi_2 \phi_2}{\sqrt{2}}&\m \phi_1 \phi_2 &\m h z\cr\vbox{\hrule}
\m w^- w^+ & 4\l_H & \sqrt{2}\l_H  & \sqrt{2}\l_H & \frac{\l_{1h}}{\sqrt{2}} & \frac{\l_{1h}}{\sqrt{2}}
    & \l_{12h} & 0                     
   \cr
\m \frac{hh}{\sqrt{2}}& \sqrt{2}\l_H & 3\l_H & \l_H & \frac{\l_{1h}}{2} & \frac{\l_{2h}}{2} & \frac{\l_{12h}}{\sqrt{2}} & 0
   \cr
\m \frac{zz}{\sqrt{2}} & \sqrt{2}\l_H & \l_H & 3\l_H & \frac{\l_{1h}}{2} & \frac{\l_{2h}}{2} & \frac{\l_{12h}}{\sqrt{2}} & 0
   \cr
\m \frac{\phi_1 \phi_1}{\sqrt{2}} & \frac{\l_{1h}}{\sqrt{2}} & \frac{\l_{1h}}{2} & \frac{\l_{1h}}{2} & \frac{\l_{1s}}{2} & \frac{\l_{e_1}}{2} & \frac{\l_{e_2}}{\sqrt{2}}  & 0
   \cr 
\m \frac{\phi_2 \phi_2}{\sqrt{2}} & \frac{\l_{2h}}{\sqrt{2}} & \frac{\l_{2h}}{2} & \frac{\l_{2h}}{2} & \frac{\l_{e_1}}{2} & \frac{\l_{2s}}{2} & \frac{\l_{e_3}}{\sqrt{2}} & 0
   \cr
\m \phi_1 \phi_2 & \l_{12h} & \frac{\l_{12h}}{\sqrt{2}} & \frac{\l_{12h}}{\sqrt{2}}& \frac{\l_{e_2}}{\sqrt{2}} & \frac{\l_{e_3}}{\sqrt{2}} & \l_{e_1} & 0  \cr
\m h z & 0 & 0 & 0 & 0 & 0 & 0 & 2 \l_H  \cr
   }\,, \\
\end{eqnarray*}
and 
\begin{eqnarray*}
\mathcal{B} &=&
  \bordermatrix{
  &\m h\phi_1||z\phi_1 &\m h\phi_2||z\phi_2 \cr\vbox{\hrule}
\m h\phi_1||z\phi_1 & \l_{1h} & \l_{12h} 
   \cr
\m h\phi_2||z\phi_2 & \l_{12h} & \l_{2h}
   \cr
   }\,.   
\end{eqnarray*} 
For singly charged particle state the  $4 \times 4$ amplitude matrix is given by
\begin{equation}
\mathcal{M}_{\rm SC}= \bordermatrix{
  &\m w^+ h &\m w^+ z &\m w^+\phi_1 &\m w^+\phi_2 \cr\vbox{\hrule}
\m w^- h & 2\l_H & 0 & 0 & 0
   \cr
   \m  w^- z & 0 & 2 \l_H & 0 & 0 \cr
\m w^- \phi_1 & 0 & 0 & \l_{1h} & \l_{12h}
   \cr
 \m w^- \phi_2 & 0 & 0 &\l_{12h} & \l_{2h}
 \cr  }
\label{eq:matrix-sc}
\end{equation}
The distinct eigenvalues of matrix  given in Eq.~\ref{eq:matrix-nc} and \ref{eq:matrix-sc} are
\begin{gather}
a = 2 \l_H  \\
b_{\pm} = \frac{1}{2}\left( \l_{1h}+\l_{2h}\pm\sqrt{4 \l^2_{12h}+\left(\l_{1h}-\l_{2h}\right)^2}\right), 
\end{gather}
and the solutions of the equation given below
\begin{equation}
\label{eq:unitarity-z2-2}
4 \,x^4 +p_1\, x^3 + q_1 \,x^2 + r_1\, x + s_1 = 0.
\end{equation}
Where, 
 \begin{align}
 \label{pqr}
 p_1 &= -2\l_{1s}-2\l_{2s}-4\l_{e_1}-24\l_{H} \, , \nonumber \\
 q_1 &=-8\l_{12h}^2-4\l_{1h}^2+\l_{1s}\l_{2s}+2\l_{1s}\l_{e_1}+12\l_{1s}\l_{H}-4\l_{2h}^2+2\l_{2s}\l_{e_1}+12\l_{2s}\l_{H}-\l_{e_1}^2\nonumber \\
 &~~+24\l_{e_1}\l_{H}-2\l_{e_2}^2-2\l_{e_3}^2\, , \nonumber \\
 r_1 &=4\lambda_{12h}^2\lambda_{1s}+4\lambda_{12h}^2\lambda_{2s}-8\lambda_{12h}\lambda_{1h}\lambda_{e_2}-8\lambda_{12h}\lambda_{2h}\lambda_{e_3}+2\lambda_{1h}^2\lambda_{2s}+4\lambda_{1h}^2\lambda_{e_1}-4\lambda_{1h}\lambda_{2h}\lambda_{e_1} \nonumber \\
 &~~+2\lambda_{1s}\lambda_{2h}^2-\lambda_{1s}\lambda_{2s}\lambda_{e_1}-6\lambda_{1s}\lambda_{2s}\lambda_{H}-12\lambda_{1s}\lambda_{e_1}\lambda_{H}+\lambda_{1s}\lambda_{e_3}^2+4\lambda_{2h}^2\lambda_{e_1}-12\lambda_{2s}\lambda_{e_1}\lambda_{H} \nonumber \\
 &~~+\lambda_{2s}\lambda_{e_2}^2+\lambda_{e_1}^3+6\lambda_{e_1}^2\lambda_{H}-2\lambda_{e_1}\lambda_{e_2}\lambda_{e_3}+12\lambda_{e_2}^2\lambda_{H}+12\lambda_{e_3}^2\lambda_{H} \, , \\
s_1 &= -2\lambda_{12h}^2\lambda_{1s}\lambda_{2s}+2\lambda_{12h}^2\lambda_{e_1}^2+4\lambda_{12h}\lambda_{1h}\lambda_{2s}\lambda_{e_2}-4\lambda_{12h}\lambda_{1h}\lambda_{e_1}\lambda_{e_3}+4\lambda_{12h}\lambda_{1s}\lambda_{2h}\lambda_{e_3}\nonumber \\ 
&~~-4\lambda_{12h}\lambda_{2h}\lambda_{e_1}\lambda_{e_2}-2\lambda_{1h}^2\lambda_{2s}\lambda_{e_1}+2\lambda_{1h}^2\lambda_{e_3}^2+4\lambda_{1h}\lambda_{2h}\lambda_{e_1}^2-4\lambda_{1h}\lambda_{2h}\lambda_{e_2}\lambda_{e_3}-2\lambda_{1s}\lambda_{2h}^2\lambda_{e_1}\nonumber \\
 &~~+6\lambda_{1s}\lambda_{2s}\lambda_{e_1}\lambda_{H}-6\lambda_{1s}\lambda_{e_3}^2\lambda_{H}+2\lambda_{2h}^2\lambda_{e_2}^2-6\lambda_{2s}\lambda_{e_2}^2\lambda_{H}-6\lambda_{e_1}^3\lambda_{H}+12\lambda_{e_1}\lambda_{e_2}\lambda_{e_3}\lambda_{H} \nonumber
\end{align}
\textbf{Minimal $\mathbb{Z}_3$ model:}\\
Identically, for the Lagrangian given in Eq. \ref{eq:mz3} there will be a $11 \times 11$ amplitude matrix for neutral particle states and $4 \times 4$ for singly charged particle states. Corresponding distinct eigenvalues are mentioned in table \ref{table:unitarity}. \\
\textbf{Two particle under $\mathbb{Z}_3$ :}\\
Similarly for multiparticle $\mathbb{Z}_3$ model we will have a $22 \times 22$ amplitude matrix for neutral particle states and a $6 \times 6$ matrix for the singly charged particle states. Constraints on the parameter space are given in  table \ref{table:unitarity}. Where $y$ is the solution of the equation given by
\begin{equation}
\label{eq:unitarity-z3-2}
y^4 + p_2 \,y^3 + q_2 \,y^2 + r_2\, y + s_2=0
\end{equation} 
Where 
\begin{align}
p_2&=-4\lambda_{1s}-4\lambda_{2s}-5\lambda_{e}-6\lambda_{H}, \nonumber \\
q_2&=-4\l_{12h}^2-2\l_{1h}^2+16\l_{1s}\l_{2s}+20\l_{1s}\l_{e}+24\l_{1s}\l_{H}-2\l_{2h}^2+20\l_{2s}\l_{e}+24\l_{2s}\l_{H}\nonumber \\ 
&~~-\l_{e}^2+30\l_{e}\l_{H},\nonumber \\
r_2&=16\l_{12h}^2\l_{1s}+16\l_{12h}^2\l_{2s}+8\l_{1h}^2\l_{2s}+10\l_{1h}^2\l_{e}-4\l_{1h}\l_{2h}\l_{e}+8\l_{1s}\l_{2h}^2-80\l_{1s}\l_{2s}\l_{e}\nonumber \\ 
&~~-96\l_{1s}\l_{2s}\l_{H}-120\l_{1s}\l_{e}\l_{H}+10\l_{2h}^2\l_{e}-120\l_{2s}\l_{e}\l_{H}+5\l_{e}^3+6\l_{e}^2\l_{H}, \\
s_2&=-64\l_{12h}^2\l_{1s}\l_{2s}+4\l_{12h}^2\l_{e}^2-40\l_{1h}^2\l_{2s}\l_{e}+20\l_{1h}\l_{2h}\l_{e}^2-40\l_{1s}\l_{2h}^2\l_{e}+480\l_{1s}\l_{2s}\l_{e}\l_{H}\nonumber \\ 
&~~-30\l_{e}^3\l_{H} \nonumber
\end{align}
\textbf{$\mathbb{Z}_3 \times \mathbb{Z}^{'}_3$ Model :}\\
Constraints on dimensionless parameters for this model has been given in table \ref{table:unitarity}. Note that $z$ is the solution of the following equation
\begin{equation}
\label{eq:unitarity-z3-z3p}
z^3 + q_3\, z^2 + r_3 \,z + s_3=0
\end{equation}
Where
\begin{align}
q_3&= -4\lambda_{1s}-4\lambda_{2s}-6\lambda_{H}, \nonumber \\
r_3&=-2\lambda_{1h}^2+16\lambda_{1s}\lambda_{2s}+24\lambda_{1s}\lambda_{H}-2\lambda_{2h}^2+24\lambda_{2s}\lambda_{H}-\lambda_{e}^2,  \\
s_3&=8\lambda_{1h}^2\lambda_{2s}-4\lambda_{1h}\lambda_{2h}\lambda_{e}+8\lambda_{1s}\lambda_{2h}^2-96\lambda_{1s}\lambda_{2s}\lambda_{H}+6\lambda_{e}^2\lambda_{H} \nonumber
\end{align}
%%%%%%%%%%%%%%%%%%   References %%%%%%%%%%%%%%%%%%%%%%%%%%%%%%%%%%%%
\bibliographystyle{JHEP}
\bibliography{z3-2-ref.bib}
%%%%%%%%%%%%%%%%%%%%%%%%%%%%%%%%%%%%%%%%%%%%%%%%%%%%%%%%%%%%%%%%%%%%

\end{document}